\documentclass[iop]{emulateapj}
\usepackage{amssymb,amsmath,graphicx}
\usepackage{adjustbox}
\usepackage{mathrsfs}
\usepackage{tabularx}
\usepackage{upgreek}

\def\hi{{\mbox{\sc Hi}}}

\def\arcdeg{\hbox{$^\circ$}}
\def\slantfrac#1#2{\hbox{$\,^#1\!/_#2$}}

\def\degree{^{\circ}}
\def\Msun{M_{\odot}}

\def\ga{\mathrel{\hbox{\rlap{\hbox{\lower4pt\hbox{$\sim$}}}\hbox{$>$}}}}
\def\la{\mathrel{\hbox{\rlap{\hbox{\lower4pt\hbox{$\sim$}}}\hbox{$<$}}}}
\newcommand\Tstrut{\rule{0pt}{2.6ex}}         
\newcommand\Bstrut{\rule[-0.9ex]{0pt}{0pt}}   

\shorttitle{Revealing the full Extent of $\hi$ around Spirals}
\shortauthors{Pingel et al.}

\begin{document}

\title{\large A GBT Survey of the HALOGAS Galaxies and Their Environments I: \\
\small Revealing the full extent of $\hi$ around NGC891, NGC925, NGC4414 \& NGC4565}
\author{N. M. Pingel\altaffilmark{1,2}} \author{D. J. Pisano\altaffilmark{1,2,3}} 
\author{G. Heald\altaffilmark{4,5}}
\author{T. H. Jarrett\altaffilmark{6}} \author{W. J. G. de Blok\altaffilmark{5,6,7}}
\author{G. I. G. J{\'o}zsa\altaffilmark{8,9,10}} 
\author{E. J\"utte\altaffilmark{12}} 
\author{R. J. Rand\altaffilmark{13}} 
\author{T. Oosterloo\altaffilmark{5,7}} \author{B. Winkel\altaffilmark{11}}
\altaffiltext{1}{Department of Physics and Astronomy, West Virginia University,
White Hall, Box 6315, Morgantown, WV 26506; nipingel@mix.wvu.edu}
\altaffiltext{2}{Center for Gravitational Waves and Cosmology, West Virginia University, 
Chestnut Ridge Research Building, Morgantown, WV 26505}
\altaffiltext{3}{Adjunct Astronomer at Green Bank Observatory, P.O. Box 2, Green Bank, WV 24944, USA.}
\altaffiltext{4}{CSIRO Astronomy and Space Science, PO Box 1130, Bentley WA 6102, Australia}
\altaffiltext{5} {Netherlands Institute for Radio Astronomy (ASTRON), Postbus 2, 7990 AA Dwingeloo, the Netherlands}
\altaffiltext{6}{Department of Astronomy, University of Cape Town, Private Bag X3, Rondebosch 7701, South Africa}
\altaffiltext{7}{Kapteyn Astronomical Institute, University of Groningen, PO Box 800, 9700 AV, Groningen, The Netherlands}
\altaffiltext{8} {South African Radio Astronomy Observatory (SARAO), SKA South Africa, The Park, Park Road, Pinelands, 7405, South Africa}
\altaffiltext{9} {Rhodes University, Radio Astronomy Research Group (RARG), PO Box 94, Grahamstown, 6140, South Africa}
\altaffiltext{10} {Argelander-Institut f\"ur Astronomie, Auf dem H\"ugel 71, 53121 Bonn, Germany}
\altaffiltext{11}{Max-Planck-Institut f{\"u}r Radioastronomie (MPIfR), Auf dem H{\"u}gel 69, 53121}
\altaffiltext{12} {Astronomisches Institut Ruhr-Universit{\"a}t Bochum, Universit{\"a}tsstra{\ss}e 150 , 44780 Bochum, Germany}
\altaffiltext{13} {Department of Physics and Astronomy, University of New Mexico, Albuquerque, NM 87131, USA}

\begin{abstract}
We present initial results from a deep neutral hydrogen ($\hi$) survey of the HALOGAS galaxy sample, which includes the spiral galaxies NGC891, NGC925, NGC4414, and NGC4565, performed with the Robert C. Byrd Green Bank Telescope (GBT). The resulting observations cover at least four deg$^2$ around these galaxies with an average 5$\sigma$ detection limit of 1.2$\times$10$^{18}$ cm$^{-2}$ over a velocity range of 20 km s$^{-1}$ and angular scale of 9.1$'$. In addition to detecting the same total flux as the GBT data, the spatial distribution of the GBT and original Westerbork Synthesis Radio Telescope (WSRT) data match well at equal spatial resolutions. The $\hi$ mass fraction below $\hi$ column densities of 10$^{19}$ cm$^{-2}$ is, on average, 2\%. We discuss the possible origins of low column density $\hi$ of nearby spiral galaxies. The absence of a considerable amount of newly detected $\hi$ by the GBT indicates these galaxies do not have significant extended diffuse $\hi$ structures, and suggests future surveys planned with the SKA and its precursors must go \textit{at least} as deep as 10$^{17}$ cm$^{-2}$ in column density to significantly increase the probability of detecting $\hi$ associated with the cosmic web and/or cold mode accretion.

\end{abstract}

\section{Introduction}\label{Intro}
\setcounter{footnote}{0}
Resolved neutral hydrogen ($\hi$) observations undertaken over the past decade have revealed many intricate details related to the morphology and dynamics of spiral galaxies. A primary science goal of recent large surveys is to develop a deep understanding of how physical processes within the disks of spiral galaxies, such as star formation and the subsequent stellar feedback, affect their local circumgalactic environments. Surveys such as The $\hi$ Nearby Galaxy Survey (THINGS; \citealt{walter08}) and Hydrogen Accretion in LOcal GAlaxies Survey (HALOGAS; \citealt{heald11}; hereby referred to as H11) performed with the Very Large Array (VLA) and Westerbork Synthesis Radio Telescope (WSRT), respectively, provide high resolution maps of the environments around nearby spiral galaxies. 

Accretion of diffuse gas onto the disks of galaxies from the intergalactic medium (IGM) is a possible explanation for how the $\hi$ content of galaxies has remained relatively constant since $z\sim$ 2 while the star formation rate was up to 10 times higher at high redshifts \citep{not12,madDickinson14}. The constant $\hi$ content implies that galaxies have somehow replenished themselves with enough gas to fuel continuous star formation. And though not directly responsible for star formation, $\hi$ is an intermediate phase towards molecular hydrogen, which is the raw ingredient of the star formation fuel. If the star formation is to continue, external gas has to be accreted and pass through the $\hi$ phase at some stage in the accretion process. Observationally inferred accretion rates as traced by $\hi$, however, fall between 0.1 and 0.2 $\Msun$  at low redshifts. This is a full order of magnitude lower than what is needed for galaxies to continually form stars at their current rates \citep{sancisi08,kauff10}. This discrepancy presents two intriguing scenarios: the cycle of star formation will eventually exhaust all of the available fuel within a few Gyr and star formation itself will gradually cease, or processes that refuel galaxies with the necessary gas have been missed by previous surveys. Numerical simulations have shown a likely mechanism for refueling star formation is through a quasi-spherical `hot' mode and filamentary  `cold' mode \citep{keres05,keres09,birnDek03}. Cold in the context of these numerical simulations refers to gas that has not been heated above the virial temperature of the galaxy's potential well ($\sim$ 10$^5$ K), and hot refers to gas that has virialized in a process akin to the classical theory of galaxy formation in which shock-heated, virialized gas with short cooling timescales accretes onto the central galaxy (e.g., \citealt{reesOs77}). These simulations also suggest cold mode accretion was the dominant form of accretion at $z \ge$ 1 for all systems, and remains prevalent through $z=0$ for galaxies in low-density environments ($n_{gal}$ $\lesssim$ 1 h$^3$ Mpc$^{-3}$) and $M_{halo}$ $\lesssim$ 10$^{11.4}$ $\Msun$ (or $M_{bary}$ $\leq$ 10$^{10.3}$ $\Msun$). For perspective, our own Milky Way has a virial (and thus halo) mass on the order of 10$^{12}$ $\Msun$. These cold flows should exist in the form of vast filaments of cold, diffuse gas that permeate through the hot halo \citep{keres05}. Comparisons by \citet{nelson13} between the smoothed particle hydrodynamic (SPH) numerical scheme employed in \citet{keres05,keres09} and more sophisticated adaptive mesh refinement (AMR) simulations revealed the relative contribution of the cold mode is likely overestimated in earlier SPH simulations due to inherent numerical deficiencies. Nevertheless, the AMR simulations do show $\textit{some}$ fraction of gas is accreted cold.

The temperature of the gas in these predicted cold filaments is too high for a significant amount of neutral gas to exist within the largely ionized medium. However, AMR hydrodynamic simulations presented by \citet{joung12} show large amplitude non-linear perturbations can create cooling instabilities in which gas is collisionally excited and cools through subsequent radiative de-excitation of excited states. Large filamentary flows of inflowing gas are a possible seeding mechanism for non-linear perturbations, which allow gas to cool enough to form $\hi$ clouds within the inner most regions of the halo (R $\leq$ 100 kpc) at $\hi$ column densities ($N_{HI}$ $\leq$ 10$^{18}$ cm$^{-2}$) currently detectable with existing telescopes. 

More recent independent ballistic models show galactic fountain activity can account for the presence of extraplanar $\hi$ around the Milky Way (in clouds like Complex C; \citealt{frat15}) and NGC891 \citep{frat06}. In addition, \citet{frat17} describes the condensation of hot coronal gas in the wake of the interaction with cooler galactic fountain gas, showing that fountain driven accretion can cool enough lower coronal gas to sufficiently extend the gas depletion time.

Observational evidence for predicted cold flows is very limited. Though absorption measurements, \citet{stocke10} and \citet{ribaudo11} both find low metallicity gas infalling onto a nearby solar metallicity Lymann Limit System whose mass is consistent with the presence of cold flows predicted by simulations. The presence of infalling, low metallicity gas is certainly consistent with cold flows, but these measurements do not reveal any information about the extended spatial distribution of the accretion. Absorption measurements are very promising in that they accurately probe the metallicity of galaxy halos, but such studies require a quasar or other bright background source to measure the absorption line of interest. Absorption studies of the Milky Way, in which these desired sightlines are abundant, show our own Galaxy is surrounded by an immense amount of low column density gas that is both ionized and neutral (e.g., \citealt{wakker03,richt17}) with temperatures ranging from 10$^2$ to 10$^7$ K. Detection in emission does not rely on serendipitous sightlines required for external galaxies, and will constrain the large scale extent of the predicted cold flows or a potential diffuse component.

The unrivaled point source response of radio interferometers like the WSRT and VLA allows for incredible high resolution mapping capabilities at angular resolutions $\sim$$\frac{\lambda}{b_{max}}$, where $b_{max}$ is the maximum baseline ($b_{max} = 2.7$ km for the WSRT), which reveals the small scale structure of galaxies. On the downside, interferometers act as spatial filters by construction, and in particular due to the minimum possible spacing between neighboring telescopes in an interferometer (i.e., the physical size of each dish), there is a gap in $\it{u-v}$ coverage at large angular scales from the absence of short baselines. This gap is often referred to as the `short-spacing' problem, and it limits the amount of large scale structure an interferometer is able to detect (e.g., \citealt{brWl85}). As a consequence of the lack of sensitivity at large angular scales, past $\hi$ observations performed with interferometers may have missed significant reservoirs of gas around galaxies. On the other hand, the $\hi$ is observed in channels covering small velocity ranges only, and thus potentially does not extend enough to cause the sampled baselines to miss several interesting low-density, diffuse features. The full $\it{u-v}$ coverage capability of single dish telescopes \citep{stan02} permits the detection of structure at all angular scales to test that notion. The unblocked aperture design of the GBT and resulting low sidelobes coupled with the compromise between resolution (9.1$'$) and high surface brightness sensitivity (T$_{sys}$ $\lesssim$ 20 K) make it the ideal instrument to look for low column density structure around the HALOGAS sources. 

The few surveys that have mapped down to $N_{HI}$ $\lesssim$ 10$^{19}$ cm$^{-2}$ have uncovered several interesting low density, diffuse features. Perhaps most notably, \citet{brth04} discovered a low column density $\hi$ filament connecting M31 and M33. Two possible explanations for its origin have been presented since its discovery: either it is similar to filaments seen in simulations of the cosmic web \citep{pop09}, and thus an observational example of the cold mode accretion process, or it was created via a past tidal interaction between M31 and M33 \citep{bek08, putman09}. Higher resolution observations with the GBT by \citet{wolfe13} and \citet{wolfe16} show that this filament is clumpy in nature and made up of small $\hi$ clouds with M$_{HI}$ $\sim$ 10$^{4-5}$ $\Msun$, $N_{HI}$ $\sim$ 10$^{18}$ cm$^{-2}$, and diameters on the order of kpc. M31 has a M$_{dyn}$ $\sim$ 1.3 $\times$10$^{12}$ $\Msun$ \citep{cor10}, which suggests the cold mode accretion scenario is unlikely. Furthermore, the total $\hi$ mass of these clouds is only 4.6$\times$10$^{6}$ $\Msun$ providing only meager neutral mass accretion rates for a conservative infall time estimates of 10$^{7-8}$ years. The origin of these clouds is still an open and intriguing question which can be answered by utilizing sensitive observations of the $\hi$ within the circumgalactic environment of M31 and M33 \citep{wolfe16}. 

Other recent detections by the GBT of large $\hi$ structures in NGC6946 by \citet{pisano14} and NGC2403 by \citet{deBlok14} suggest these features are seen around a variety of galaxies. In order to determine the true origin of these filaments, resolve the discrepancy between observed accretion rates and SFRs, and obtain a comprehensive understanding of how the disks of galaxies interact with their surrounding circumgalactic environment, a comprehensive $\hi$ census spanning a wide range of astrophysically interesting properties (e.g., dynamical mass, total $\hi$ mass, halo mass, SFRs, etc) is required. A complete census of these properties will build up large number statistics and uncover any underlying correlations between intrinsic galaxy properties and possible signatures of accretion. The HALOGAS observations of 24 nearby galaxies obtained with the WSRT and the THINGS survey with the VLA are critical steps towards just such a census. To ensure this census is absolutely complete, interferometer observations must be supplemented with large single dish observations to cover all angular scales to assure large-scale emission is not resolved out by interferometers, and to map down to the lowest possible column density levels.

In this pilot paper we present data and analysis from four sources from the HALOGAS survey: NGC891, NGC925, NGC4414, and NGC4565. These GBT maps are among the deepest ($N_{HI}$$\sim$10$^{18}$ cm$^{-2}$) for external galaxies obtained to date in $\hi$. This paper serves as an introduction to the full survey as a way to outline our analysis methods and highlight our overall goals. In Section~\ref{section:Sample} we present an overview of the HALOGAS sample; the observing configuration, reduction strategy and a discussion on our GBT beam model and how we convolve the WSRT data to avoid contamination from extended structure are outlined in Section~\ref{section:ObsRed}. The results from our comparison between the GBT and WSRT data for our initial four sources are discussed in Section~\ref{section:Results} with an investigation into how the diffuse $\hi$ environment relate to intrinsic galaxy properties following in Section~\ref{section:Discussion}. We then summarize our conclusions and commenting on future work in Section~\ref{section:conclusion}.

\section{The HALOGAS Data and Sample}\label{section:Sample}
\subsection{WSRT Data Cubes}

The high-resolution HALOGAS cubes were produced from data obtained with the WSRT. See H11 for complete details on the observational configuration and data reduction of these data; see also \citet{oost07} for configuration and reduction details specific to NGC891. A particular aspect of the HALOGAS observational setup we wish to highlight here is the minimum baseline length of 36 m, which translates to a maximum recoverable angular scale of $\sim$20$'$. As mentioned in the Introduction, this particular angular scale is important because smooth emission extending above this limit will not be present in the WSRT data but fully observable by the GBT.

As will be discussed below, a significant portion of our analysis relies on convolved WSRT data convolved to the GBT resolution. Flux measurements from convolved data have the potential to be misleading as convolved emission will extend outside the original `clean' boundaries used to produce the final interferometer cubes. Fully cleaned maps of array data are the sum of the restored clean components and residual map. Generally, only a portion of a map is cleaned, and thus, will have the correct flux. The flux in uncleaned portions of the map will be overestimated by a factor equal to the ratio of the dirty beam size to clean beam size. We measure outside of the clean region because the convolution obviously extends source emission beyond its original boundaries. Including regions of uncleaned emission in the convolution will inevitably lead to misleading total flux measurements as pixels with uncleaned emission have an intensity scale defined as Jy per dirty beam as opposed to Jy per clean beam (e.g., \citealt{jorMor95}). In the interest of quantifying how the inclusion of unclean emission will affect the overall flux measurements in the convolved WSRT data, we extract two sub-maps from the original WSRT high-resolution NGC891 cubes that include a different number of pixels that were excluded from the cleaning as part of the data reduction. We then compare the total $\hi$ flux values after convolving these sub-maps as described in Section~\ref{subsection:convSec}. Specifically, $\sim$60\% of the first extracted subregion includes pixels that were part of the original clean region, while the remaining 40\% of the pixels in this sub-map were excluded from the cleaning algorithm. In the second sub-map, $\sim$90\% of the pixels were cleaned during data reduction, while the remaining 10\% of the pixels were excluded from cleaning. We find the total flux measurements of these two sub-maps to be the same. This indicates that while some uncleaned emission --- whether from pixels not inside the original clean region or simply low-level emission below the original clean threshold --- is inevitably included, the total flux estimates are not significantly affected. That said, there are still systematic calibration uncertainties introduced by the specific treatment of the raw $\it{u-v}$ visibilities to consider. A few examples are the removal of residual baseline structure, flagged/missing baselines, antenna shadowing, and different weighting schemes applied to the complex visibilities. We adopt an overall systematic flux uncertainty of 5\% to encapsulate uncertainties related to how the WSRT data were processed. 

\subsection{Sample}
The total HALOGAS sample consists of 24 unbarred and barred, nearby spirals that span a very diverse range of astrophysically significant properties such as star formation rates (SFRs), $\hi$ mass (M$_{HI}$), stellar mass (M$_{*}$), baryonic mass (M$_{bary}$), etc. The sample also consists of galaxies with a wide range of environments. 

We adopt the best distance values listed for each source in H11. Given that the~`best' distances were determined by taking the median measurements from well established methods (e.g., Cepheid and/or tip of the red giant branch, planetary-nebulae luminosity function, Tully-Fisher distances), we adopt a conservative 10\% overall uncertainty for these distances. The SFRs for all galaxies mentioned in this work besides NGC4565 and NGC2997 (to be discussed in Section~\ref{subsection:cmaRelation} in regards to similar GBT data of nearby galaxies) are computed utilizing data from the 22$\mu$m band of the Wide-field Infrared Survey Explorer (WISE), a space based observatory deployed to map the entire sky in the infrared, along with far-ultraviolet (FUV) luminosity data from the Galaxy Evolution Explorer (GALEX; \citealt{gil07}). We follow the method outlined in \citet{jarrett13} to obtain an infrared SFR (SFR$_{IR}$) that traces star formation obscured by dust with further calibrations derived by \citet{cluver14} for the same IR bands. Due to the difficulty of disentangling the relative contribution from young and old stellar populations to various Polycyclic aromatic hydrocarbon emission bands near 12$\mu$m, we chose to use the SFR$_{IR}$ derived from the WISE 22$\mu$m band. The SFR tracing UV photons associated with young massive stars is given as 
\begin{equation}\label{eq:SFR_FUV}
log_{10}\left(\frac{SFR_{FUV}}{\Msun\;\textrm{yr$^{-1}$}}\right) = log_{10}\left(\frac{L_{FUV}}{L_{\odot}}\right)-9.69,
\end{equation}
and is derived from the calibrations of \citet{buat08,buat11}. \citet{jarrett13} combine $SFR_{IR}$ and $SFR_{FUV}$ to estimate a total SFR using the form 
\begin{equation}
SFR_{tot} = (1-\eta)SFR_{IR} + SFR_{FUV}, 
\end{equation}
where $\eta$ represents the fractional contribution to the total IR emission from dust reradiating energy injected from old stars; the value of 0.17$\pm$0.1 is adopted \citep{buat11,jarrett13}. In the case of NGC4565, the SFR is taken from the HALOGAS calculations in \citet{heald12} since no FUV luminosities were available for these sources through $GALEX$. 

The 3.4 $\mu$m band of $WISE$ effectively traces light from old stars resulting in a practical measure of the stellar mass. \citet{jarrett13} show a linear trend exists between $WISE$ W1$-$W2 and W2$-$W3 color and stellar mass-to-light ratios ($M/L$). By relating the WISE color to 2MASS $K_s$ in-band luminosity, a $M/L$ ratio  can be derived from the $K_s$ stellar mass relation of \citet{zhu10}. This trend is further explored in \citet{cluver14}, in which the W1$-$W2 color and stellar masses from the Galaxy And Mass Assembly (GAMA; \citealt{driver09,driver11}) survey are used for empirical calibration of the relationship. The best-fit for their sample including both passive and star-forming systems, but excluding known active galactic nuclei sources and WISE colors dominated by nuclear activity (W1$-$W2 $\ge$ 0.8), is
\begin{equation}
log_{10}\left(\frac{M_{*}/\Msun}{L_{W1}/L_{\odot}}\right) = -1.96\left(W_{3.4\mu m} - W_{4.6\mu m}\right)-0.03.
\end{equation}
We adopt this relation to determine aggregate stellar masses for the HALOGAS sources. See \citet{jarrett13} and \citet{cluver14} for explicit details on the calculating of aggregate stellar masses utilizing $WISE$ data. The adopted distance and SFRs from H11 are summarized in Table~\ref{tab:configParams}. See Table 2 in H11 for a complete summary of the targets' properties and explanation for how certain target properties such as SFRs and distances were derived. 

\section{GBT Observations and Data Reduction and Low-Resolution WSRT Cubes}\label{section:ObsRed}
\subsection{Observations and Data Reduction}\label{subsection:ObsRedGBT}

\begin{table*}
\resizebox{\textwidth}{!}{\begin{tabular}{lcccccccc}
    \hline \hline
     \\[-1.0em]
    Source & $\alpha$\tablenotemark{a} & $\delta$\tablenotemark{b} & Systemic Velocity [km s$^{-1}$] & Total Bandwidth [MHz] & $\Delta$$v$[km s$^{-1}$]\tablenotemark{c} & $\sigma$ [mK]  \tablenotemark{d} & $d_{best}$ [Mpc]  \tablenotemark{e}& SFR$_{tot}$ [$\Msun$ yr$^{-1}$]  \tablenotemark{f} \\
     \\[-1.0em]
     \hline
    NGC891 & 02$^h$22$^m$33.4$^s$ & 42$^{\circ}$20$'$57$''$  & 528$\pm$4 &  50.0  & 5.15 & 10 & 9.2$\pm$0.9 & 3.92$\pm$1.75 \\
    NGC925  & 02$^h$27$^m$16.9$^s$ & 33$^{\circ}$34$'$45$''$  & 553$\pm$3 & 50.0  &  5.15 & 12   & 9.1$\pm$0.9 & 0.91$\pm$0.16 \\ 
    NGC4414 & 12$^h$26$^m$27.1$^s$ & 31$^{\circ}$13$'$25$''$  & 716$\pm$6 &  23.4  & 5.15 & 13 &  18$\pm$2  & 3.45$\pm$1.58  \\
    NGC4565 & 12$^h$36$^m$20.8$^s$ & 25$^{\circ}$59$'$16$''$  & 1230$\pm$5  & 50.0 &  5.15  & 15 & 11$\pm$1  & 0.67$\pm$0.10 \\
    \\[-1.0em] 
    \hline
       \end{tabular}}
   \caption {Summary of Observations and Properties}
\tablenotetext{1}{Right Ascension (J2000)}
\tablenotetext{2}{Declination (J2000)}
\tablenotetext{3}{Velocity Resolution}
\tablenotetext{4}{final rms noise per velocity channel}
\tablenotetext{5}{adopted distance from H11}
\tablenotetext{6}{SFRs derived from WISE; value for NGC4565 taken from \citep{heald12}}
\label{tab:configParams} 
\end{table*}

Our GBT maps were made in a ``basket-weave'' fashion by scanning the telescope for 2$^{\circ}$/3$^{\circ}$ along constant lines of Right Ascension ($\alpha$$_{J2000}$) and Declination ($\delta$$_{J2000}$) to stitch together a final 4 deg$^2$ (9 deg$^2$ for NGC925) image \citep{mang07}. If potential cold flows exist, their visibility in $\hi$ depends on how close to the disk the relatively warm gas of the flow transitions to the $\hi$ phase. For the range of distances of the sources presented in this work, the angular span of 2 deg corresponds to approximately 315 kpc -- 620 kpc, which are sufficient to capture a majority of the virial volume. Each row or column is offset by 3$'$, and each scan consisted of a total of 72 separate 5 s integrations that were dumped every 100$''$ to ensure Nyquist sampling. NGC891 and NGC925 were observed during January 2010 as part of the GBT project 10A-026, while NGC4565 and NGC4414 were observed during October 2013 and 2014 as part of GBT projects 13B-406 and 14B-293, respectively. We obtained an additional ten hours to map the inner 2x1 deg$^2$ region of NGC891 as part of GBT project 16A-411. NGC925 and NGC4565 were observed with the GBT Spectrometer as the backend while the Versatile GBT Astronomical Spectrometer (VEGAS) was the backend used for the observations of NGC4414. The observations of NGC891 we present in this work combine the initial data from the GBT Spectrometer with the additional ten hours of data which utilized VEGAS as the backend. During each observation, the band was centered on the $\hi$ line at the redshifts of the sources. The observing bandwidth, frequency resolution, noise at the native velocity resolution, and other data properties are summarized in Table~\ref{tab:configParams}. Calibration during the observation was done by frequency switching $-$30.0\% ($-$23.4\% for observations with VEGAS as the backend) of the total bandwidth from the center frequency at a one second period, and each observation session included time on 3C48, 3C147, or 3C295 as primary flux calibrator in order to compute a T$_{cal}$ value for the noise diode. T$_{cal}$ values computed for the Spectrometer varied between 1.53 and 1.57 K for both XX and YY polarizations with remarkably low scatter on the mean values of 1.56 and 1.57 for over a period of several months (1$\sigma$ $\sim$ 0.01 K). The exceptional stability in the T$_{cal}$ values translate to an upper limit on the uncertainty in the relative flux calibration to $<$ 1\% (more on the absolute calibration below). The computed T$_{cal}$ values for observations performed with VEGAS as the backend showed a significant decrease in the stability. We compute the mean T$_{cal}$ for VEGAS-only observations to be 1.41 K and 1.43 K for the YY and XX polarizations, respectively with a $1\sigma$ uncertainty equal to 0.05 K for each. The increase scatter in the VEGAS T$_{cal}$ was likely due to a crosstalk issue in the electronics of the backend that pushed the upper limit on the absolute flux calibration slightly upwards to 5\%, including other systematics such as baseline removal and, to a lesser extent, T$_{sys}$ variation. The crosstalk issue manifested itself by offseting the relative power levels between the XX and YY polarizations. This issue was fixed by the time we obtained our new data on NGC891 and consequently only NGC4414 data are affected. To ensure the derived T$_{cal}$ values for VEGAS did not affect our flux calibration, we observed each HALOGAS source that used VEGAS as the backend in the Fall of 2016 as part of GBT project GBT16B-393. These observations were done explicitly in position-switching mode where the GBT was centered on source for five minutes, and then moved two degrees in Right Ascension to obtain a five minute off spectrum. Comparing spatially overlapping integrations between our original VEGAS data and those from the deep pointing observations showed very good agreement within the noise when the polarizations are averaged to derive the Stokes I component. We are therefore confident that the data presented here are not significantly affected by the large variation in the T$_{cal}$ values. We still adopt an overall 5\% flux uncertainty as an upper limit to account for possible errors in bandpass calibration, interference, and errors in modeling atmospheric effects. Given that the WSRT observations largely used the same calibrator sources, we are also confident in the relative flux calibrations between the two data sets. We determine the gain to be 1.86 K Jy$^{-1}$ based on the computed T$_{cal}$ values and an $\sim$0.65 aperture efficiency at 1420 MHz \citep{booth11}. The typical system temperature of observations fell between 15 K and 20 K.

The frequency switched data were reduced using a custom GBTIDL\footnote{http://gbtidl.nrao.edu/} routine to calculate a source temperature of the form
\begin{equation}
T_{src} = T_{CAL}\cdot\frac{P_{s}-\langle P_{s,off}\rangle}{\langle P_{s,off}^\mathrm{CalOn} - P_{s,off}^\mathrm{CalOff}\rangle},
\label{eq:calib}
\end{equation} 
where the quantities in brackets denote averages of four integrations from each edge of the map for a total of eight `off' integrations to increase the signal-to-noise. $P_s$ refers to average power of the CalOn (noise diode on) and CalOff (noise diode off) states --- i.e., (CalOn+CalOff)/2 --- for a single integration in either the reference ($s=0$) or signal ($s=1$) bandpass switching state. $P_{off}$ is therefore the arithmetic mean of the CalOn and CalOff states for the `off' spectrum generated from the average of the eight edge integrations in the corresponding bandpass state. Additionally, the denominator represents the average difference between CalOn and CalOff states for the corresponding switching bandpass. Equation~\ref{eq:calib} therefore calibrates our frequency-switched data as if we had position switched data. Using the edge of the map as a reference position is an advantage because we are able to obtain a reference spectrum for each RA/Dec scan without sacrificing telescope time to slew off source. In the case of NGC925, emission from a companion source near the edge of our map forced us to use eight contiguous integrations on the opposite map edge as our `off' position. The reference bandpass ($s=0$) is then shifted in frequency to match the signal  bandpass ($s=1$) and arithmetically averaged to improve the noise by a factor of $\sqrt{2}$. We then fit a third order polynomial to the emission free regions of the spectra for all of our sources to remove baseline structure and any residual continuum sources leftover from our calibration procedure. In order to convert to units of $T^{*}_{A}$ we assumed a constant zenith opacity appropriate for 21cm observations of 0.01 \citep{chyn08}. We manually flagged < 0.5\% of all integrations due to broadband radio frequency interference using a custom graphical flagging GBTIDL routine. Finally, we used a boxcar smoothing function to produce raw GBT spectra at a velocity resolution of 5.15 km s$^{-1}$. See \citet{wolfe16} and the GBT technical memo \citet{wolfeMemo15} for a more comprehensive discussion on the mapping sensitivity, data acquisition, and reduction. 

These calibrated data were then converted to a format readable by AIPS using the IDLTOSDFITS\footnote{Developed by Glen Langston of NRAO; documentation at https://safe.nrao.edu/wiki/bin/view/GB/Data/IdlToSdfits.} program and imported into AIPS where they were gridded into the final raw GBT data cubes using a convolution function of a Gaussian-tapered circular Bessel function \citep{mang07} within the task SDGRD.

\subsection{Convolution of WSRT data}\label{subsection:convSec}

 Accurate comparisons between high-resolution interferometer observations and single dish observations require the interferometer data to be smoothed to the resolution of the single-dish data set. Conventional analysis approximates the single-dish beam with a Gaussian smoothing function. While this is normally a good approximation, our goal of detecting low column density material around nearby galaxies requires a more comprehensive treatment since the Gaussian approximation for the single-dish beam does not take into account radiation coming into the near sidelobes and wings of the main beam. Since our sources are well separated from Galactic $\hi$ we are only concerned with emission originating from $\hi$ in and near our target galaxies. 

\begin{figure*}
\centering{\includegraphics[width=4.0in]{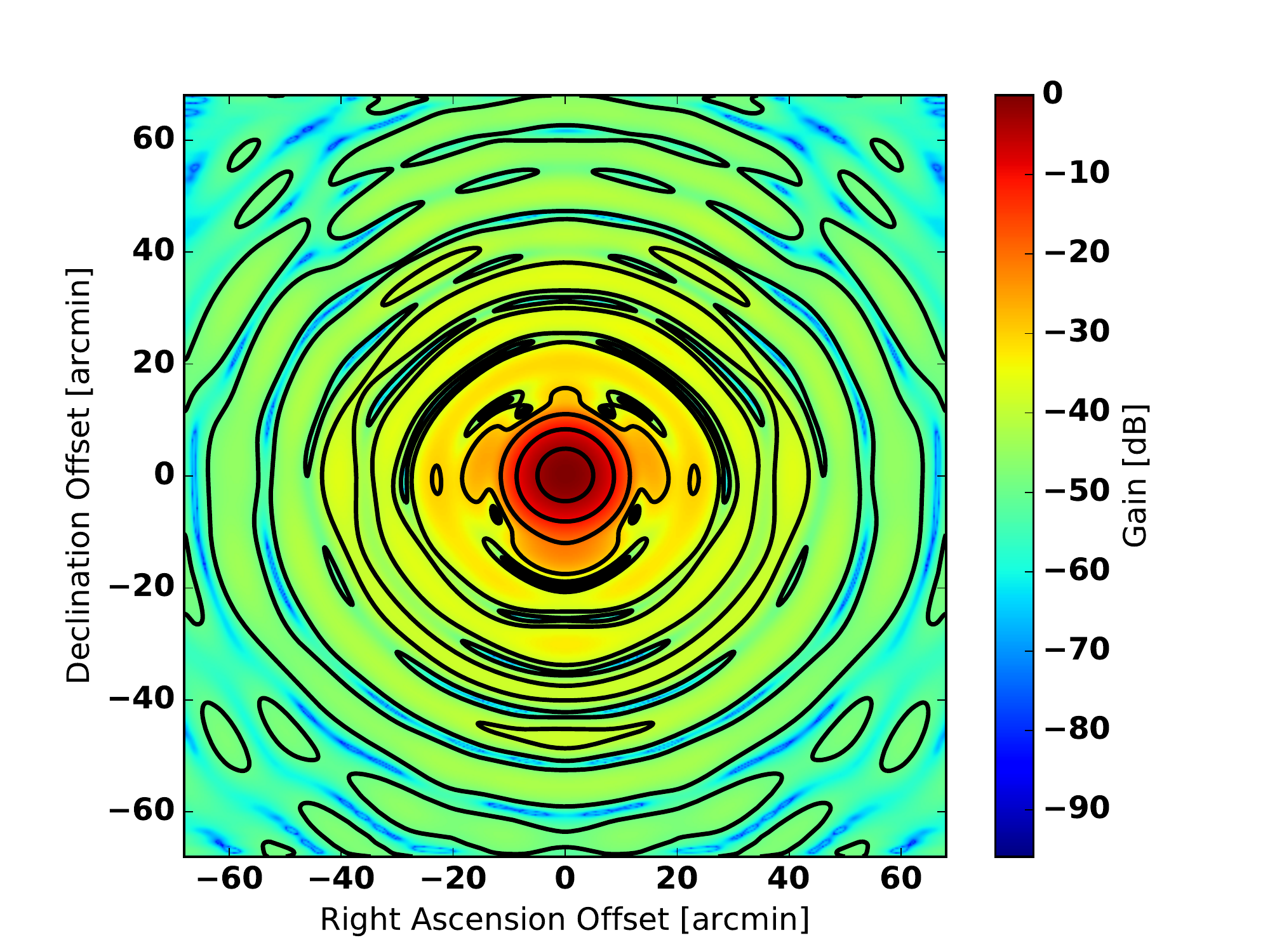}}
\caption{Normalized polar plot of a model GBT beam response on the 
sky. The contours represent dB levels of -3, -10, -20, -30, and -40.}
\label{fig:beamMaps}
\end{figure*}

In order to account for extended emission coming into the GBT's innermost sidelobes, we construct a model beammap motivated from recent detailed simulations of the aperture illumination of the GBT at 1.4 GHz (Srikanth 2017; private commmunication) as part of an ongoing project to produce a measured map of the GBT power pattern. We derive the model by taking the square of the forward Fourier Transform of a simulated in-focus aperture radiation pattern. An example of our calculated beam model is shown in Figure~\ref{fig:beamMaps} as a contoured 2D polar plot. The farthest sidelobes in this map extend out approximately 1.2$\degree$ on the sky in radius between the $-$40 and $-$50 dB level. Interestingly, the innermost sidelobes $\textit{are not}$ symmetric. \citet{booth11} has thoroughly mapped the all-sky response of the GBT beam out at the farthest sidelobes. But due to system saturation effects within 1$\degree$ of the Sun, the nearest sidelobes, which are most susceptible to contamination from stray radiation in this work, were not able to be accurately mapped. The model presented in that paper (see their Figure 5) showed the sidelobes were highly symmetric. Emission entering the sidelobes at $\sim$1.2$\degree$ are not likely to contaminate our spectra as the gain here is approximately 50 dB down from the peak of the primary beam.  However, the first sidelobe is only 30 dB down from the peak response. This makes contamination possible with some sources that have strong, extended emission out to $\sim$0.5$\degree$. We account for the possibility of contamination in the nearest sidelobes by using the GBT model beam map to include the sidelobe geometry in the convolution of the WSRT cubes. This provides the most accurate comparison possible between the two data sets without a genuine, measured beam map of the GBT. 

There are, however, subtle consequences from utilizing a beam model with a set size as a smoothing kernel for high-resolution data. Namely, the final angular resolution of the convolved data will be \textit{slightly} coarser than the GBT data because the high-resolution data are already convolved with the WSRT clean beam. Since the GBT beam model is not strictly Gaussian, we cannot solve for the optimal size of the kernel by taking the quadrature difference between the respective full-width half maximum (FWHM) values of the GBT and WSRT clean beam. We instead employ a Fourier Transform ($\mathfrak{F}$) deconvolution method wherein the $\mathfrak{F}$ of the GBT beam model ($\mathfrak{F}\left[\Omega_{GBT}\right]$) is divided by the $\mathfrak{F}$ of the WSRT clean beam ($\mathfrak{F}\left[\Omega_{WSRT}\right]$) to derive an optimal smoothing kernel, $K\left\{WSRT\rightarrow GBT\right\}$.  

We follow a similar procedure described in \citet{aniano11} to apply the necessary tapers that reduce numerical noise introduced by the FFT algorithm. We first taper the FT of the GBT beam with a piecewise function of the form

\[
\Phi\left(k\right)=
\begin{cases}
1 &\text{if } k < k_{\alpha} \\
exp\left[-(1.8249\times\left(\frac{k-k_{\alpha}}{k_{\beta}-k_{\alpha}} \right)\right]&\text{if } k_{\alpha} \leq k \leq k_{\beta} \\
0 &\text{if }  k_{\beta} < k \\
\end{cases}
\]

where $k_{\beta}$ is taken to be the spatial frequency corresponding to four times the FWHM of the GBT and $k_{\alpha} = 0.9k_{\beta}$. We refer to the tapered form of $\mathfrak{F}\left[\Omega_{GBT}\right]$ as $\mathfrak{F}\left[\Omega_{GBT}\right]_{\Phi}$. The amount of power removed from tapering the high spatial frequencies is much less than 1\% since almost all of the power is contained near spatial frequencies corresponding to the main beam. Nevertheless, it is important to ensure high spatial frequencies are near zero to avoid introducing numerical artifacts when $\mathfrak{F}\left[\Omega_{WSRT}\right]$ is divided out. We then take the reciprocal of $\mathfrak{F}\left[\Omega_{WSRT}\right]$ and taper the result by the low-pass filter
\[
f_{T}\left(k\right)=
\begin{cases}
1 &\text{if } k < k_{L} \\
0.5\times\left[1+\text{cos}\left(\pi\times\frac{k-k_{L}}{k_{H}-k_{L}} \right)\right]&\text{if } k_{L} \leq k \leq k_{H} \\
0 &\text{if }  k_{H} < k. \\
\end{cases}
\]
Here $k_{H}$ is chosen such that $\mathfrak{F}\left[\Omega_{WSRT}\right]\left(k_H\right) = 5\times10^{-3}\cdot max \left(\mathfrak{F}\left[\Omega_{WSRT}\right]\right)$ and $k_L$ = 0.7$\times$k$_H$. We choose $k_H$ as such to ensure a high spatial frequency cutoff where $\mathfrak{F}\left[\Omega_{WSRT}\right]$ is still appreciable, while the form of $k_L$ is chosen to leave a majority of the lowest spatial frequency components unaffected by the filter.

The form of the optimal kernel is therefore given as
\begin{equation}
K\left\{WSRT\rightarrow GBT\right\} =\left\lvert \mathfrak{F}^{-1}\left[\mathfrak{F}\left[\Omega_{GBT}\right]_{\Phi}\times\frac{f_T}{\mathfrak{F}\left[\Omega_{WSRT}\right]}\right] \right\rvert,
\label{eq:kernel}
\end{equation}
where $\left\lvert\mathfrak{F}^{-1}\right\rvert$ represents the magnitude of the inverse Fourier Transform back to the sky plane. We note that \citet{aniano11} worked strictly with the real components since their kernels were largely rotationally symmetric. Since the WSRT clean beams are generally not symmetric, their Fourier Transforms will not be rotationally symmetric. As such, we work with the polar forms of $\mathfrak{F}\left[\Omega_{WSRT}\right]$ and $\mathfrak{F}\left[\Omega_{GBT}\right]$ to preserve the phase contribution. The resulting smoothing kernels are, as expected, \textit{marginally} narrower than the GBT beam model with the largest residuals ($\leq$ 0.01\%) occurring towards the center. 

\subsection{Primary-beam Correction}\label{subsection:PBC}
The standard reduction techniques of imaging and deconvolution of interferometer data result in a model representation of the sky multiplied by the primary-beam response of the antennas. The most accurate measure of flux requires a `primary-beam correction' to the final data products, which we define as dividing out the primary-beam response of each velocity plane in the cube. We remove the primary-beam response from the native high-resolution WSRT cubes in the $Miriad$ software package \citep{sault95} with the same beam model used in H11. The primary-beam correction is applied before convolution as this most accurately represents the sky distribution observed by the interferometer. Furthermore, since the resulting low-resolution data cube must be scaled by the ratio of the GBT beam to the smaller WSRT synthesized beam to account for resolution differences when measuring the total flux, we found that removing the primary beam response after convolution does not conserve the total flux value computed for the high-resolution cube. To avoid issues with the non-uniform noise properties towards the edge, we extract a sub-cube such that all spatial scales fall within the half-power point of the WSRT primary beam for our subsequent analysis.

\section{Results}\label{section:Results}
\subsection{Summary of Analysis}
\label{analysisSummary}
\begin{table*}
\centering{\begin{tabular}{lcccccc}
    \hline \hline
     \\[-1.0em]
    Source & $\theta_{maj}$ [arcseconds] & $\theta_{min}$ [arcseconds] & $\sigma$ [K] & S [Jy km s$^{-1}$] & 3$\sigma$ $N_{HI}$ [cm$^{-2}$] & $\Delta$v [km s$^{-1}$] \\
\hline \\[-0.75em]
\textbf{NGC891} & & & & & \\
High-Res WSRT & 28.0 & 21.4 & 0.100 & (1.90$\pm$0.09)$\times$10$^{2}$ & 6.7$\times$10$^{19}$ & 8.24 \\ 
Conv. WSRT & 546.0 & 546.0 & 0.003 & (1.91$\pm$0.09)$\times$10$^{2}$ & 2.0$\times$10$^{18}$ & 8.24 \\ 
Regridded GBT & 546.0 & 546.0 & 0.010 & (1.93$\pm$0.09)$\times$10$^{2}$ & 6.8$\times$10$^{18}$ & 8.24 \\
\hline \\[-0.75em]
\textbf{NGC925} & & & & & \\
High-Res WSRT & 37.9 & 33.2 & 0.130 & (2.85$\pm$0.14)$\times$10$^{2}$ & 2.4$\times$10$^{19}$ & 4.12 \\ 
Conv. WSRT & 546.0 & 546.0 & 0.003 & (2.84$\pm$0.14)$\times$10$^{2}$ & 1.8$\times$10$^{18}$ & 4.12 \\
Regridded GBT & 546.0 & 546.0 & 0.013 & (2.96$\pm$0.15)$\times$10$^{2}$ & 4.0$\times$10$^{18}$ & 4.12 \\
 \hline \\[-0.75em]
\textbf{NGC4414} & & & & & & \\
High-Res WSRT & 39.0 & 33.5 & 0.130 & (6.2$\pm$0.3)$\times$10$^{1}$ & 5.2$\times$10$^{19}$ & 4.12 \\ 
Conv. WSRT & 546.0 & 546.0 & 0.005 & (6.1$\pm$0.3)$\times$10$^{1}$ & 2.0$\times$10$^{18}$ & 4.12 \\
Regridded GBT & 546.0 & 546.0 & 0.013 & (7.3$\pm$0.4)$\times$10$^{1}$ & 5.2$\times$10$^{18}$ & 4.12 \\
\hline \\[-0.75em]
\textbf{NGC4565} & & & & & & \\
High-Res WSRT & 33.5 & 30.8 & 0.150 & (2.74$\pm$0.14)$\times$10$^{2}$ & 4.8$\times$10$^{19}$ & 4.12 \\ 
Conv. WSRT & 546.0 & 546.0 & 0.004 & (2.71$\pm$0.14)$\times$10$^{2}$ & 1.7$\times$10$^{18}$ & 4.12 \\
Regridded GBT & 546.0 & 546.0 & 0.015 & (2.66$\pm$0.13)$\times$10$^{2}$ & 7.0$\times$10$^{18}$ & 4.12 \\

\hline
\end{tabular}}
\caption {Summary of Data Cubes}
\label{tab:mapPropSummary} 
\end{table*}

In the following section, we present initial analysis for four (NGC891, NGC925, NGC4414, and NGC4565) of the 24 total sources in the HALOGAS sample, By analyzing $\hi$ at various angular resolution and sensitivities, the extent of $\hi$ environment of these galaxies between 18 $\lesssim$ $log_{10}$($N_{HI}$/cm$^{2}$) $\lesssim$ 21 can be fully characterized. Before discussing results for individual sources, we first summarize the steps of our analysis.

\subsubsection{Global $\hi$ Profiles and Noise}
\label{subsubsec:HIProfiles}
The flux as a function of velocity measured in the three data cubes is shown in each target's respective subsection. We first use the $Miriad$ task REGRID to regrid each GBT cube to be on the same angular and spectral scale as their WSRT counterpart. We estimate the noise properties within each respective cube, $\sigma\left(x,y\right)$, by fitting the negative half of a histogram whose pixels values were drawn from a region with no emission in all velocity channels (i.e., those with and without emission) with a Gaussian. We determine the 1$\sigma$ noise in the native, primary-beam corrected high-resolution WSRT data cubes to be between 100 and 150 mK, between 10 mK and 15 mK for the regridded GBT cubes, and between 5 and 10 mK for the convolved, primary-beam corrected WSRT cubes. These noise properties are summarized in the fourth column of Table~\ref{tab:mapPropSummary}

\subsubsection{$N_{HI}$ Images}
Since we are mostly interested in the low column density environments of our sources, care must be taken to correctly scale the convolved data, distinguish signal from noise, as well as calculating associated uncertainties. We do this for the cumulative $\hi$ mass as a function of radius and $N_{HI}$.

To this end, we determine the gain to convert the convolved, primary-beam corrected WSRT cubes from Jy/Beam to brightness temperature in units of Kelvin using the equation
\begin{equation}
T_b = \frac{\lambda^2S}{2k\Omega_a}.
\label{eq:gainWithoutUnits}
\end{equation}
Here $S$ is the flux density, $\Omega_a$ is the beam solid angle, $k$ is the Boltzmann constant and $\lambda$ is the wavelength of the observation (i.e., 0.211 meters). Taking these values and simplifying we arrive at
\begin{equation}
T_b \left[\textrm{K}\right] = \frac{6.87\times10^{5} S \left[\textrm{Jy/beam}\right]}{\Omega_a \left[\textrm{arcseconds}^2\right]}.
\label{eq:gainWithUnits}
\end{equation}
The area of the GBT beam model used in this study is 3.69$\times$10$^{5}$ square arcseconds, while the area of the WSRT clean beam can be approximated as a Gaussian and is given by 1.1331$\times$$\left(\theta_{maj}\cdot\theta_{min}\right)$ (the major and minor axis in arcseconds, respectively). If one multiplies Equation~\ref{eq:gainWithUnits} 
by the reciprocal of of the flux density, $S$, and plugs in the area of the GBT beam model, it returns the gain factor of 1.86 [K/Jy] as derived in Section~\ref{subsection:ObsRedGBT}.

Computing the noise on an individual pixel basis is imperative to the treatment of the WSRT data (at both high and low resolutions) since the primary-beam correction changes the behavior of the noise as a function of position. The noise in the GBT data is relatively uniform over the cube, though the characterization of individual pixel noise is useful for constraining uncertainties in subsequent analysis. Since the original WSRT cubes were Hanning smoothed to their final velocity resolution, the pixels along the velocity axis are not independent.~\citet{verh01} show an associated 1$\sigma$ noise map can be computed by 
\begin{equation}
\sigma_{N}(x,y) = \sqrt{\left(N\left(x,y\right)-\frac{3}{4}\right)}\frac{4}{\sqrt{6}}\sigma\left(x,y\right), 
\label{eq:1sigNoise}
\end{equation}
where $N(x,y)$ is the number of pixels used in the integration. For each data cube we produce masked (where emission below some threshold is blanked) and unmasked $N_{HI}$ images. Masked images are useful when studying the radial dependence on column density since we wish to characterize the spatial variations of low-level signal, while unmasked images are used when studying the properties of the total flux.

We follow \citet{verh01} and \citet{lelli14} by first constructing a mask for our high-resolution data by spatially smoothing the high-resolution cubes to 40$''$ (50$''$ in the case of NGC4414) and only include pixels above 3$\sigma\left(x,y\right)$ --- as determined from a fit to the negative half of a histogram --- in the sum. As a consequence of this mask application, the number of channels used in the sum, and thus the uncertainty, will vary pixel-to-pixel in the resulting 1$\sigma$ noise maps. Due to the variation across the map, a global 3$\sigma$-level is no longer straightforward to calculate. We calculate a global 3$\sigma$ $N_{HI}$-level in these cubes by creating a signal-to-noise (S/N) map by taking the ratio of the masked $N_{HI}$ images with the 1$\sigma$ noise maps. We then take the average of pixels in the S/N maps satisfying 2.75 $\leq$ S/N $\leq$ 3.25 to ensure a large enough sample to compute a reliable mean value. The final $N_{HI}$ images only contain pixels whose value lies above this 3$\sigma$-level. In the case of the low-resolution WSRT and GBT data, we repeat the masking procedure as described above but directly discard pixels which do not meet the 3$\sigma\left(x,y\right)$ threshold without constructing a spatially coarser cube. The beam sizes, $\sigma\left(x,y\right)$ values, $\sigma\left(x,y\right)$, total flux, the $N_{HI}$ 3$\sigma$-level, and velocity resolutions for the data sets are summarized in Table~\ref{tab:mapPropSummary}.

\subsubsection{The Cumulative $\hi$ Mass vs. $N_{HI}$}
\label{subsubsec:CumulativeHIMass}
Since $\hi$ mass is simply proportional to column density times a physical area, and we have the distance to each source, we can convert an individual pixel value of column density to an equivalent $\hi$ mass to determine the $\hi$ mass probability distribution function, which measures the total $\hi$ contained within discrete $N_{HI}$ bins. Integrating this distribution therefore gives the \textit{cumulative $\hi$ mass as function of $N_{HI}$}, which measures the total $\hi$ mass for pixels equal to or exceeding $N_{HI}$ bins. This distribution conveniently describes the fraction of $\hi$ mass above and below distinct $N_{HI}$ thresholds. We can use the cumulative $\hi$ mass distribution as a diagnostic for how well the WSRT data recover the extended $\hi$ around these sources. For example, if the GBT data detected an extended diffuse $\hi$ feature that was resolved out by the WSRT, the cumulative $\hi$ mass distribution should deviate at lower column densities. In all cases presented here, the profiles flatten out well before the $N_{HI}$ 3$\sigma$-level listed in Table~\ref{tab:mapPropSummary}. We set the lowest bin to be equal to one half the listed $N_{HI}$ value in order to avoid the inclusion of noisy pixels and focus on the behavior between low and intermediate $N_{HI}$ levels. The maximum bin size is set to 90\% of the peak $N_{HI}$ value. Note that the cumulative $\hi$ mass as a function of $N_{HI}$ from both telescopes are normalized by the total GBT $\hi$ mass.

One aspect of concern when analyzing low-resolution data is whether emission adequately fills the larger GBT beam. In order to simulate the effect of this analysis on an unresolved source --- where the low resolution map would simply trace the response of the GBT beam --- we scale the GBT beam model introduced in Section~\ref{subsection:convSec} to the peak column density value of the GBT map and repeat the analysis calculating the cumulative $\hi$ mass as a function of $N_{HI}$ (we do the same for the radial profiles of the of $N_{HI}$ discussed in the next subsection). While the resolution effects will be source dependent, it is generally true that the large GBT beam does not hinder this (nor the radial average of $N_{HI}$) analysis until the highest $N_{HI}$ bins, which trace the structure of the main lobe.

\subsubsection{Radial Profiles of $N_{HI}$ and Cumulative Flux}
The radial functions of the mean column density and cumulative flux are useful to compare how the properties of the $\hi$ emission detected by the GBT and WSRT change at various angular extents. For example, a large positive offset of cumulative flux and mean $N_{HI}$ at large radii in the GBT data would indicate the WSRT resolved out large-scale $\hi$ emission. Additionally, profiles of the cumulative flux as a function of radius that do not flatten out or begin to dip at large radii may trace artifacts (e.g., negative bowls) in the high-resolution WSRT data. In the analysis of the radial extent of low column density structure, we use the masked $N_{HI}$ images to ensure low-level signal has not been buried in the noise. The cumulative flux as a function of radius is derived from unmasked maps to best probe the radial variations in total flux. We also note that by measuring properties contained within radial bins overlaid on non-axisymmetric structure, the deviations at large radii inherently only reflect the distribution of $\hi$ along the major axis; thus, these plots do not reveal where potential differences between the maps occur. Nevertheless, any potential excess $\hi$ emission originating from structures spanning larger angular scales than what the WSRT is sensitive to should be evident even in radially averaged quantities. Finally, we note that no correction for inclination angle has been applied in the computation of these radial profiles. While the inclination of our sources range between $\sim$50 deg to $\sim$90 deg, we are most interested in the relative difference between the profiles derived between the GBT and WSRT data sets rather than how the radial properties change source-to-source. This relative offset between data sets for a particular source will be the unaffected by the correction factor of the cosine of the inclination angle.

As a further check of resolution effects, the solid black line in the radial column density profiles again denote the GBT beam model scaled to the peak $N_{HI}$ value of the GBT data. Again, the model of an unresolved source deviates sufficiently well from the data indicating this analysis is not particularly hindered by resolution effects.



\subsection{NGC891}\label{subsection:n891}
\begin{figure*}
\includegraphics[width=7 in]{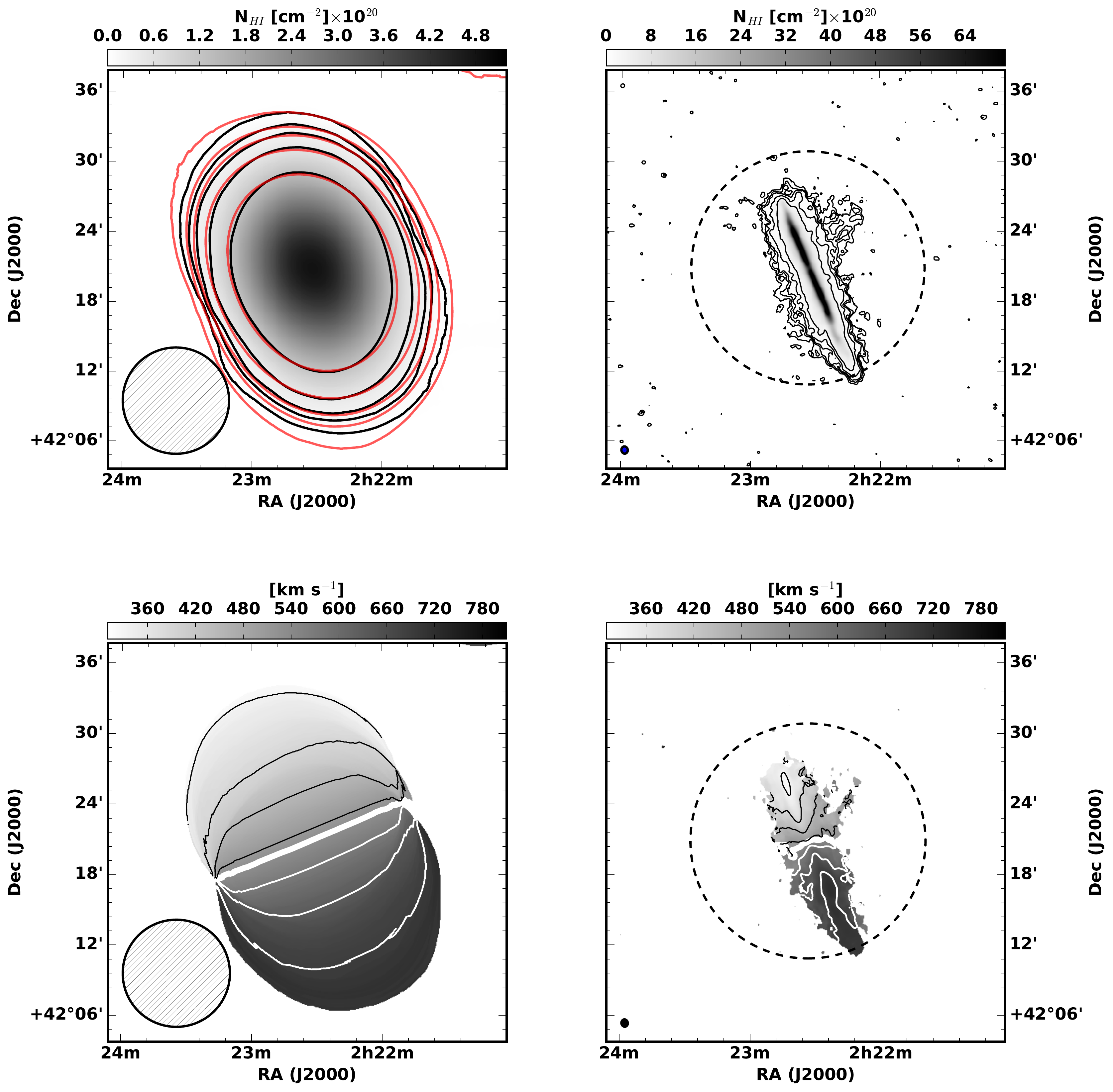}
\caption{Integrated $\hi$ image (top) and velocity fields (bottom) for the low-resolution (left) and high-resolution WSRT (right) data for NGC891. The contours in the low-resolution $N_{HI}$ images start at a column density value of 5$\times$10$^{18}$ cm$^{-2}$ and continue at 3, 5, 10, and 25 times that level. The black and red contours respectively denote the regridded GBT and convolved WSRT data. The contours in the associated high-resolution image begin at a level equivalent to 2$\times$10$^{19}$ cm$^{-2}$ and continue at 3, 5, 10, and 25 times that level. Note that the low-resolution velocity field is derived only from the GBT data cube. The contours in both velocity fields begin at 330 km s$^{-1}$ and continue in steps of 60 km s$^{-1}$. The systemic velocity of 528 km s$^{-1}$ is represented by the thick line, and the approaching and receding velocities are denoted by black and white contours, respectively. The dashed circles in the right two panels represent the maximum recoverable angular scale of the WSRT data.}
\label{fig:n891ContAll}
\end{figure*}

NGC891 is an edge on ($i$ $\gtrsim$ 89$\arcdeg$) Sb/SBb galaxy whose $\hi$ has been extensively studied over the past two decades (e.g., \citealt{san79, rup91, swa97, oost07}). Utilizing the WSRT, \citet{oost07} made deep, high-resolution maps of NGC891. These deep maps revealed a huge galactic halo containing 30\% of the total $\hi$ mass. Other significant results from this study were: the discovery of a filament extending 22 kpc in projection from the disk towards the companion, UGC1807; counterrotating $\hi$ clouds in the halo; and differential rotation lagging with respect to the disk. Models of interactions between material from galactic fountain activity and hot coronal gas have successfully recreated the observed velocity rotational velocity gradient \citep{marinacci10, frat17}. Halo pressure gradients and magnetic tension likely contribute as well~\citep{ben02}. See Figure~\ref{fig:n891ContAll} for the masked integrated $\hi$ image and velocity fields of NGC891 at low and high resolution. The GBT data do not reveal any low level emission extending from NGC891 to its companion, UGC1807.

\begin{figure*}
\centering{\includegraphics[width=7.5 in]{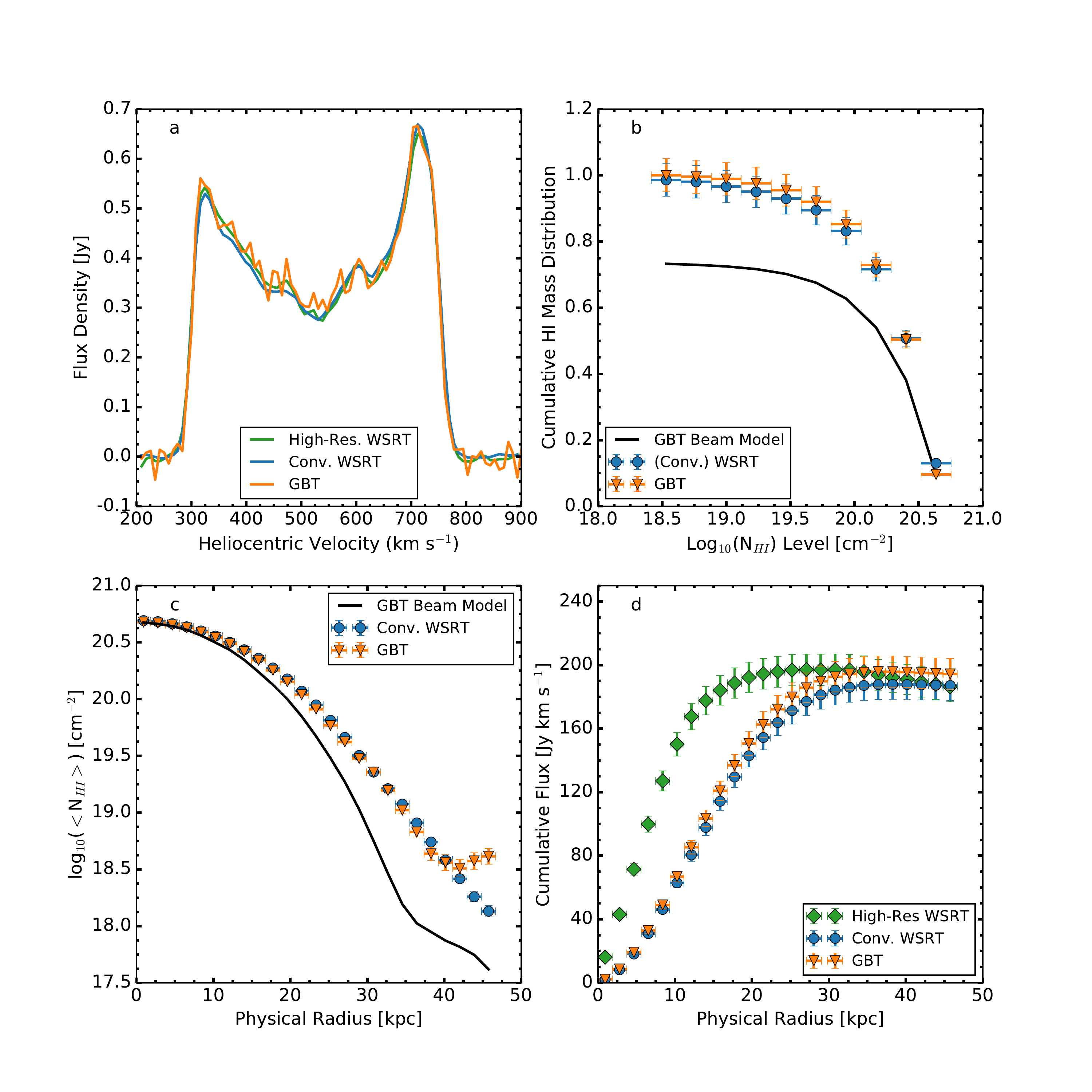}}
\caption{Comparison between the high-resolution (green diamonds) WSRT, convolved WSRT (blue circles), and  regridded GBT (orange inverted triangles) data sets of NGC891. $\textit{a}$: global $\hi$ profile; $\textit{b}$: Cumulative $\hi$ mass as a function of $N_{HI}$. The solid black line simulates an unresolved observation with our GBT beam model; $\textit{c}$: projected physical radial dependence on the azimuthally averaged $N_{HI}$; $\textit{d}$: projected radial dependence of the cumulative flux. In this case, we also show the results of our analysis on the high-resolution WSRT data.}
\label{fig:n891_4PanelSummary}
\end{figure*}

The global flux density profiles for NGC891 derived from the GBT and WSRT data are shown in Figure~\ref{fig:n891_4PanelSummary}a. There is excellent consistency between the profiles. The total measured flux for the GBT data is 193$\pm$9 Jy km s$^{-1}$. The flux density measured by the GBT translates to a total $\hi$ mass of (3.86$\pm$0.19$)\times$10$^{9}$ $\Msun$ at the adopted distance of 9.2$\pm$0.9 Mpc from H11. The flux values for the  convolved and high-resolution WSRT data are measured respectively to be 191$\pm$9 Jy km s$^{-1}$ and 190$\pm$9 Jy km s$^{-1}$.

Figure~\ref{fig:n891_4PanelSummary}b summarizes the results of the cumulative $\hi$ mass as a function of $N_{HI}$ for NGC891. While the cumulated $\hi$ mass distribution from the simulated point source observation falls well below the data, values computed from the unmasked GBT and convolved WSRT $N_{HI}$ images trace each other extremely well within the estimated uncertainties.

Figure~\ref{fig:n891_4PanelSummary}c and d show the radial dependence of the mean column density and cumulative flux, respectively. The radial dependence of column density in the GBT and convolved WSRT data are effectively identical within 40 kpc, indicating the GBT data do not reveal any large-scale $\hi$ features (down to the $\hi$ column density sensitivity limit) that may potentially related to the substantial extraplanar $\hi$ component and filament observed in NGC891. In Figure~\ref{fig:n891_4PanelSummary}d, the additional flux detected in the high-resolution WSRT data at projected scales smaller than the GBT beam (about 24 kpc at a distance of 9.2 Mpc) originates from emission that completely fills the smaller WSRT beam while remaining unresolved in the larger GBT beam. Past this point, all three data sets begin to converge to a similar value and profile shape. The dip in cumulative flux in the high-resolution profile highlights the presence of artifacts (e.g., a large-scale negative bowl caused by missing central baselines in the $\it{u-v}$ coverage, or residual sidelobes leftover from the deconvolution of the dirty beam).

\begin{figure*}
\includegraphics[width=7 in]{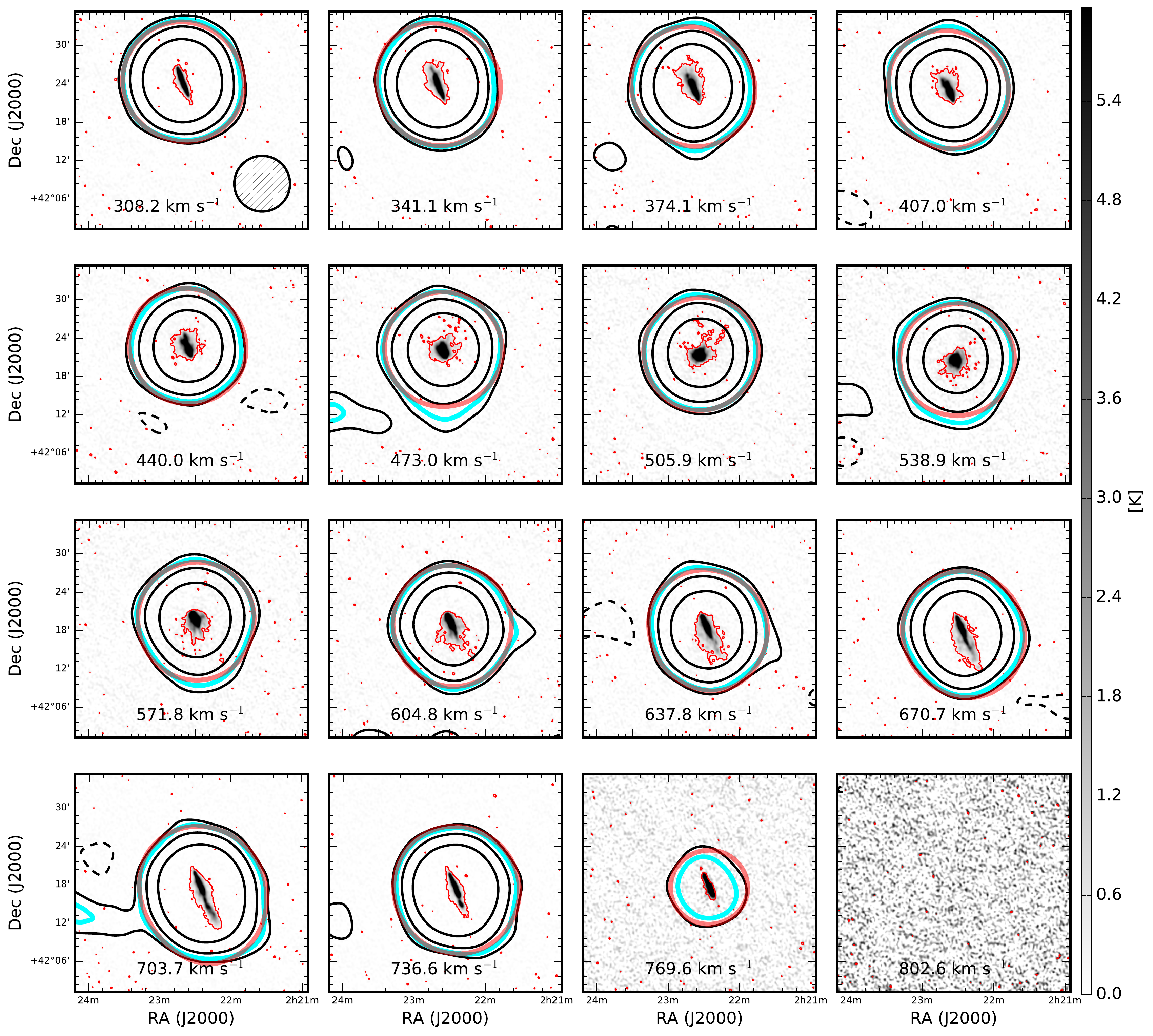}
\caption{Selected channel maps of the NGC891 high-resolution WSRT data cube with corresponding GBT and convolved WSRT contours superimposed. The GBT data are shown in black and cyan contours at levels of -3 (dashed), 3, 5 (thick cyan), and 25 times 0.01 K, or equivalently  a column density level of 1.5$\times$10$^{17}$ cm$^{-2}$ per 8.24 km s$^{-1}$ velocity channel. The grey-scale shows the $\hi$ emission from the WSRT cube. The thin red line  denotes a brightness temperature of 0.4 K, or a column density level of 5.9$\times$10$^{18}$ cm$^{-2}$, and the thick red line denotes emission at 7.5$\times$10$^{17}$ cm$^{-2}$ (the same level of the cyan contour) in the primary-beam corrected WSRT cube convolved down to the GBT resolution. The GBT beam is shown in the top left panel.}
\label{fig:NGC891FullChanMaps}
\end{figure*}

The channel maps in Figure~\ref{fig:NGC891FullChanMaps} show the extent of the emission detected by the GBT at the 5$\sigma$ level traces the same 5$\sigma$ level in the WSRT data very well. Unlike similar GBT observations of NGC2403 \citep{deBlok14}, another galaxy with a large $\hi$ filament, we do not detect any obvious structure associated with the 22 kpc long filament detected in the deep $\hi$ images presented by \citet{oost07} in the individual channel maps. The agreement between the various data sets for NGC891 in total flux, mean column density as a function of radius, and the spatial extent of emission in the individual low-resolution data channel maps show the HALOGAS data do an excellent job at recovering the full $\hi$ distribution of NGC891; there is little difference between the $\hi$ environment at the $N_{HI}$$\sim$10$^{18}$cm$^{-2}$ as compared to the $N_{HI}$$\sim$10$^{19}$cm$^{-2}$ level.

\subsection{NGC925}\label{subsection:n925}
\begin{figure*}
\includegraphics[width=7 in]{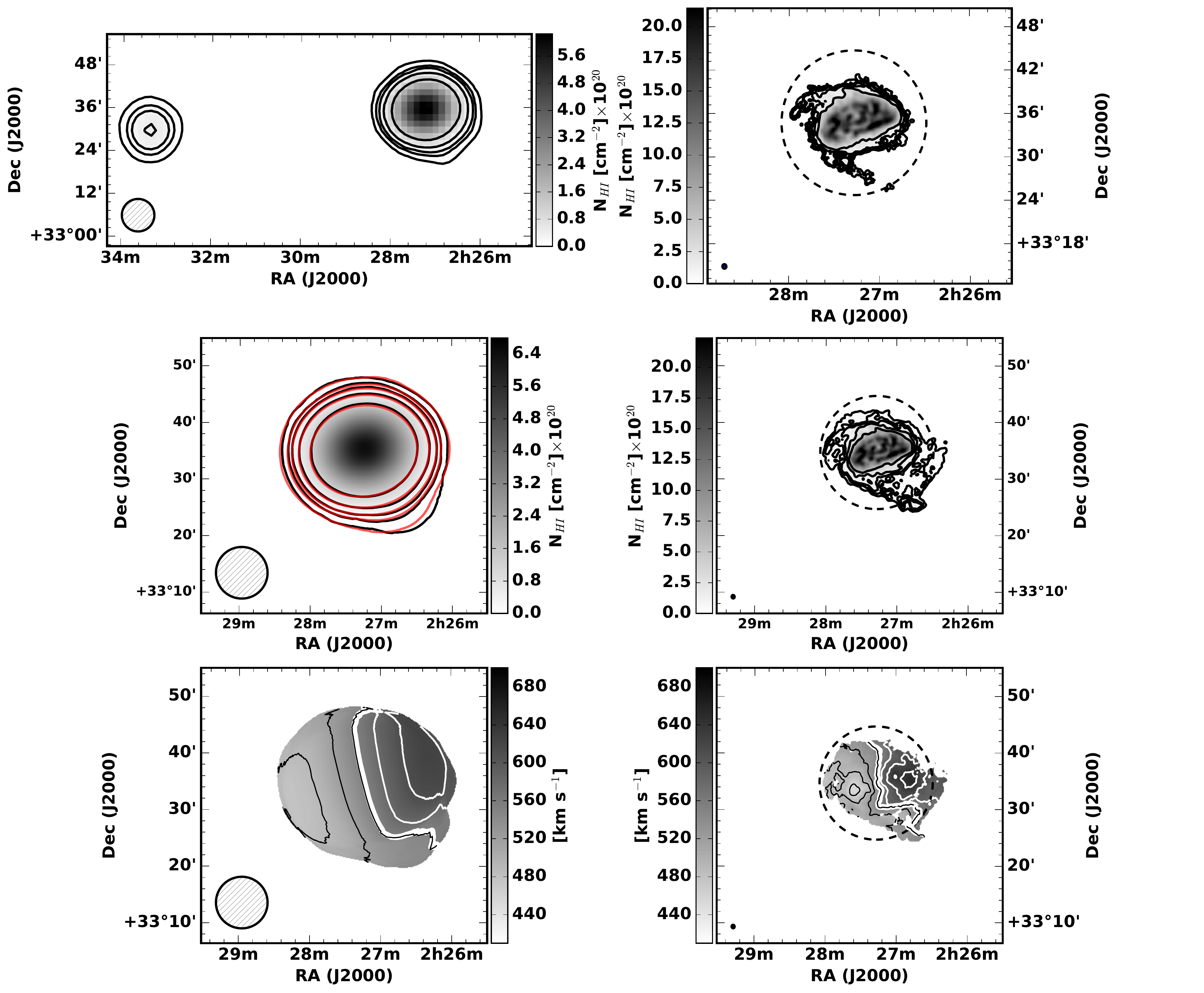}
\caption{A comparison of several data sets for NGC925. The top left panel shows an integrated $\hi$ map of the GBT data for NGC925. UGC2023, a companion to the east of NGC925, is shown in the GBT map to highlight the absence a diffuse $\hi$ bridge connecting the two galaxies. The top right panel shows an $N_{HI}$ image computed from VLA-THINGS data. The middle and bottom left panels show a zoomed $N_{HI}$ image of the low resolution cubes data and GBT velocity field. The middle and bottom right panels show the $N_{HI}$ image derived from the WSRT data and associated velocity field. The red and black contours in the low resolution $N_{HI}$ images respectively denote WSRT and GBT data; they begin at column density values of 5.0$\times$10$^{18}$ cm$^{-2}$ and continue at 3, 5, 10, and 25 times that level, while the contour levels in the VLA and WSRT $N_{HI}$ images begin at a $N_{HI}$ level equivalent to 2.0$\times$10$^{19}$ cm$^{-2}$ and continue at 3, 5, 10, and 25 times that level.  The velocity field contours begin at 430 km s$^{-1}$ and continue in steps of 30 km s$^{-1}$. The systemic velocity of 553 km s$^{-1}$ is represented by the thick white line, and the approaching and receding velocities. The dashed circles in the right-hand panels represent the maximum recoverable angular scales of the WSRT and VLA data.}
\label{fig:n925ContAll} 
\end{figure*}
 
NGC925 is another galaxy within the HALOGAS survey whose $\hi$ distribution has been thoroughly studied over past decades (e.g., \citealt{gott80, pisano98, walter08}; H11). For a galaxy such as NGC925, which has been observed as part of both THINGS and HALOGAS, we can compare the VLA, WSRT, and GBT observations. For this comparison, we utilize the same naturally weighted, residual scaled, blanked cube used to measure the total $\hi$ flux and moment maps in \citet{walter08} regridded to the WSRT spatial/spectral scale and convolved to the same angular resolution. Figure~\ref{fig:n925ContAll} shows two $N_{HI}$ images (top and middle left) of NGC925 derived from GBT data. The expanded map shows the companion galaxy, UGC2023. There is no evidence for interacting or connecting material between NGC925 and UGC2023 seen in the GBT data. Figure~\ref{fig:n925ContAll} also shows a comparison between the WSRT-HALOGAS (middle right) and VLA-THINGS (top right) data for NGC925 in the form of $N_{HI}$ images. It is clear that the more sensitive WSRT observations reveal a much more extended $\hi$ distribution than the THINGS data. This large structure $\sim$26 kpc across, which extends from $\alpha_{J2000}$=02$^{h}$24$^{m}$30$^{s}$, $\delta_{J2000}$=33$^{\circ}$17$'$ to $\alpha_{J2000}$=02$^{h}$23$^{m}$48$^{s}$, $\delta_{J2000}$=33$^{\circ}$12$'$, is visible in the VLA THINGS data, albeit at very low levels. As the shortest baselines in the THINGS survey are effectively equal to those in HALOGAS at 35 m, we can attribute the additional structure observed by the WSRT strictly to a lower noise floor (as opposed to differences in the maximum recoverable angular scale). Figure~\ref{fig:n925ContAll} also shows the velocity field maps computed from the GBT (bottom left) and WSRT (bottom right) data with the contour levels are given in the Figure caption. The high resolution velocity field shows deviations from axial symmetry coincident the the disturbed structure in the $N_{HI}$, indicating a deviation from circular rotation. H11 attributes the origin to a possible interaction with a gas-rich dwarf companion seen as a faint enhancement in the Digital Sky Survey optical plates centered at about $\alpha$$_{J2000}$=02$^{h}$26$^{m}$44$^{s}$, $\delta$=33$^{\circ}$25$'$20$''$.

\begin{figure*}
\centering{\includegraphics[width=7.5 in]{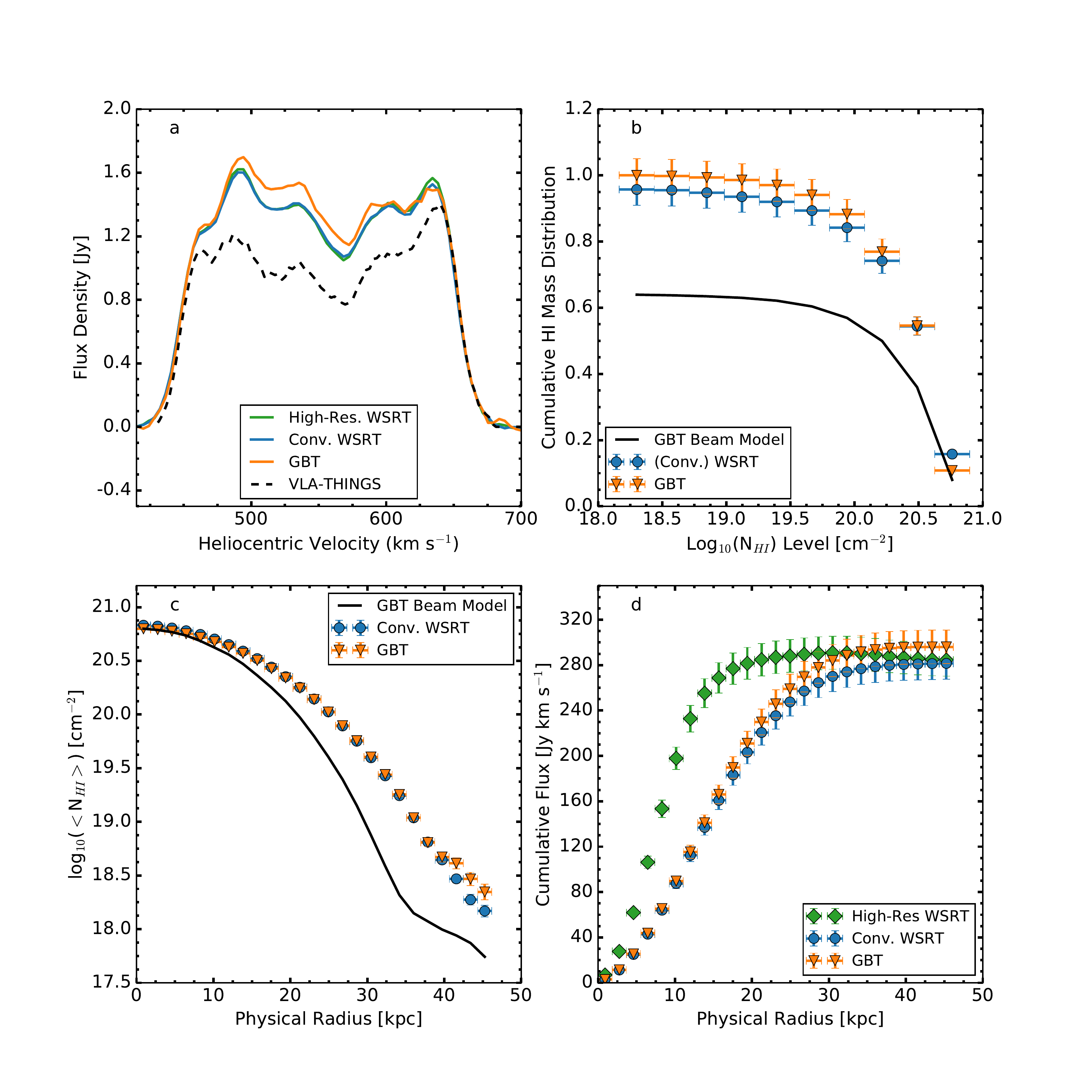}}
\caption{Comparison between the high-resolution (green diamonds) WSRT, convolved WSRT (blue circles), and  regridded GBT (orange inverted triangles) data sets of NGC925. $\textit{a}$: global $\hi$ profile; $\textit{b}$: Cumulative $\hi$ mass as a function of $N_{HI}$. The dashed and solid black lines simulate the contribution of a Gaussian beam and our GBT beam model, respectively; $\textit{c}$: projected physical radial dependence on the azimuthally averaged $N_{HI}$; $\textit{d}$: projected radial dependence of the cumulative flux. In this case, we also show the results of our analysis on the high-resolution WSRT data.}
\label{fig:n925_4PanelSummary}
\end{figure*}

We show the global HI profiles of NGC925 computed over the same angular area from the GBT, VLA, and the two WSRT data sets in Figure~\ref{fig:n925_4PanelSummary}a. The GBT detects more flux over approaching velocities (approximately 430 km s$^{-1}$ to 550 km s$^{-1}$) than the WSRT, while both the WSRT and GBT detect more flux over the entire velocity range than what is measured in VLA data.

The difference in the profiles highlights that the recovered flux detected by an interferometer is dependent on the distribution and treatment (e.g., tapering of baseline amplitudes) of the complex visibilities. For this specific THINGS cube, the visibilities were `naturally' weighted in the AIPS task IMAGR, meaning the visibilities were weighted to maximize surface brightness sensitivity. In the case of NGC925, the WSRT data were designed to maximize both sensitivity (natural -- same weight) and control over the dirty beam (uniform -- density) with the robust parameter set to 0 in the $Miriad$ task, INVERT. An additional 30$''$ Gaussian taper was also applied to the higher spatial frequencies to further maximize sensitivity to faint extended emission. It is therefore a testament to the WSRT observations that approximately 20\% more $\hi$ is detected as compared to the VLA data considering both were optimized to observe extended structure. In addition to the differences in weighting schemes, the antenna positions between the VLA and WSRT cause immutable differences in $\it{u-v}$ coverage making direct comparisons in terms of the total flux between these data sets impossible. We can therefore only conclude that, because of the agreement between both high and low-resolution WSRT and GBT flux profiles (which contains the zero-spacings information), there is excellent recovery of the diffuse $\hi$ in the WSRT-HALOGAS data for NGC925.

The cumulative $\hi$ mass as a function of $N_{HI}$ for NGC925 is presented in Figure~\ref{fig:n925_4PanelSummary}b. Just as is the case with NGC891, the cumulative $\hi$ mass as a function of $N_{HI}$ measured by the GBT is traced reasonably well by the convolved WSRT data with only slight hints of excess $\hi$ in the $N_{HI}$ bins below $log_{10}\left(N_{HI}/cm^{2}\right)$ = 20.5. The radial profiles of mean column density and cumulative flux also show consistency within the calculated uncertainties and expected behavior between the high resolution and convolved WSRT data. The slight decrease in the cumulative flux beginning at a projected physical radii 40 kpc also indicates the presence of a negative bowl in the WSRT data.
 
\begin{figure*}
\includegraphics[width=7 in]{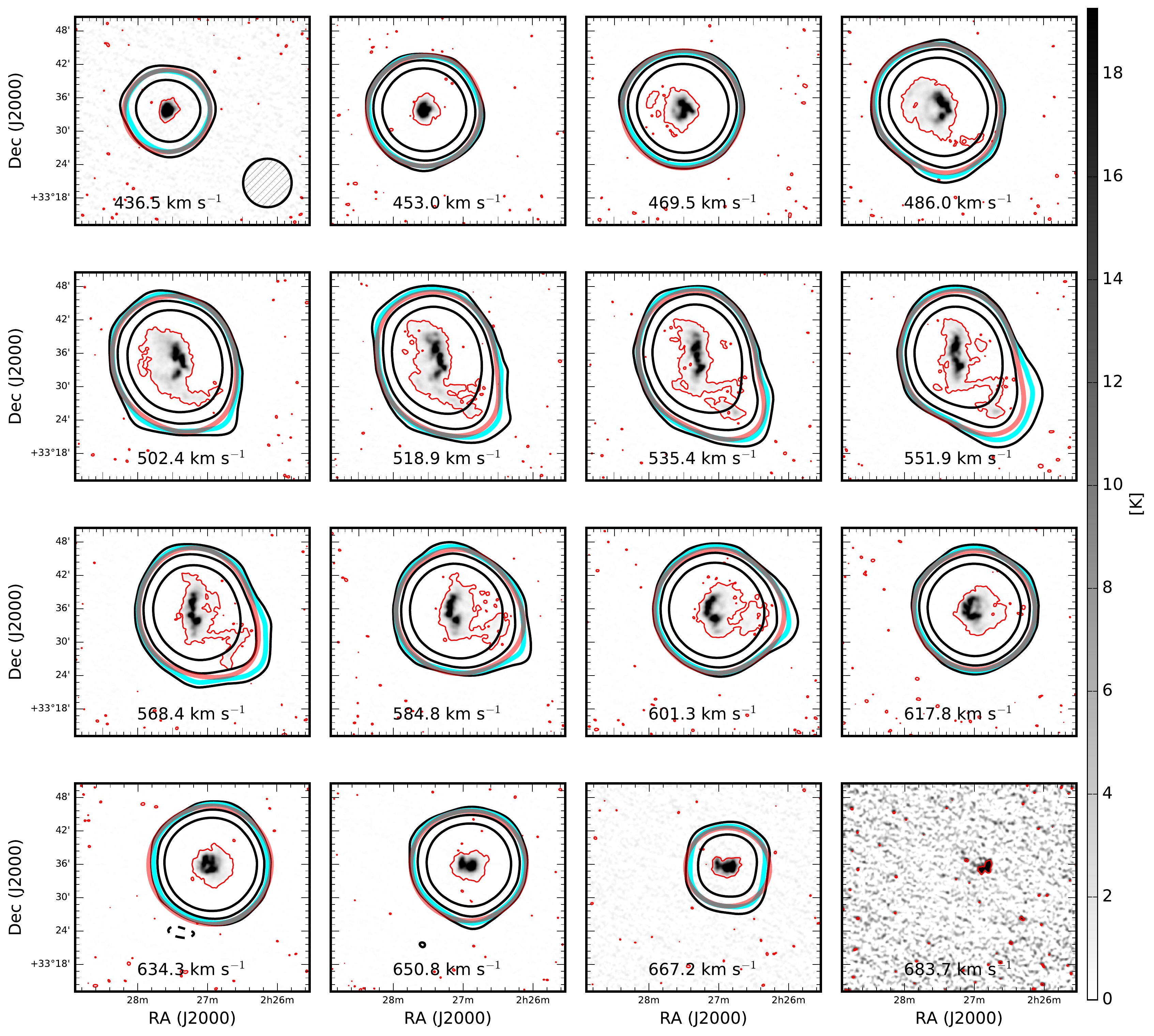}
\caption{Selected channel maps of the NGC925 WSRT data cube with corresponding GBT channel maps superimposed. The GBT data are shown in black and cyan contours at levels of -3 (dashed), 3, 5 (thick cyan), and 25 times 0.01 K, or equivalently a column density level of 9.75$\times$10$^{16}$ cm$^{-2}$ per 4.12 km s$^{-1}$ velocity channel. The grey-scale shows the $\hi$ emission from the WSRT cube. The thin red line  denotes a brightness temperature of 0.36 K, or a column density levels at 2.70$\times$10$^{18}$ cm$^{-2}$, and the thick red line denotes emission at 4.88$\times$10$^{18}$ cm$^{-2}$ (the same level of the cyan contour) in the primary-beam corrected WSRT cube convolved down to the GBT resolution. The GBT beam is shown in the top left panel.}
\label{fig:NGC925FullChanMaps}
\end{figure*}

Selected channel maps of NGC925 from the WSRT-HALOGAS data are shown in greyscale in Figure~\ref{fig:NGC925FullChanMaps} with corresponding contours denoting emission from the GBT and convolved, primary-beam WSRT data overlaid. The GBT shows $\textit{slightly}$ more extended emission in some velocity channels. A majority of the slight extension corresponds to the velocity range of 500-600 km s$^{-1}$ where the disturbed $\hi$ distribution seen in the high-resolution $N_{HI}$ images of Figures~\ref{fig:n925ContAll} is most prevalent.



\begin{figure}
\includegraphics[width = 3.5 in]{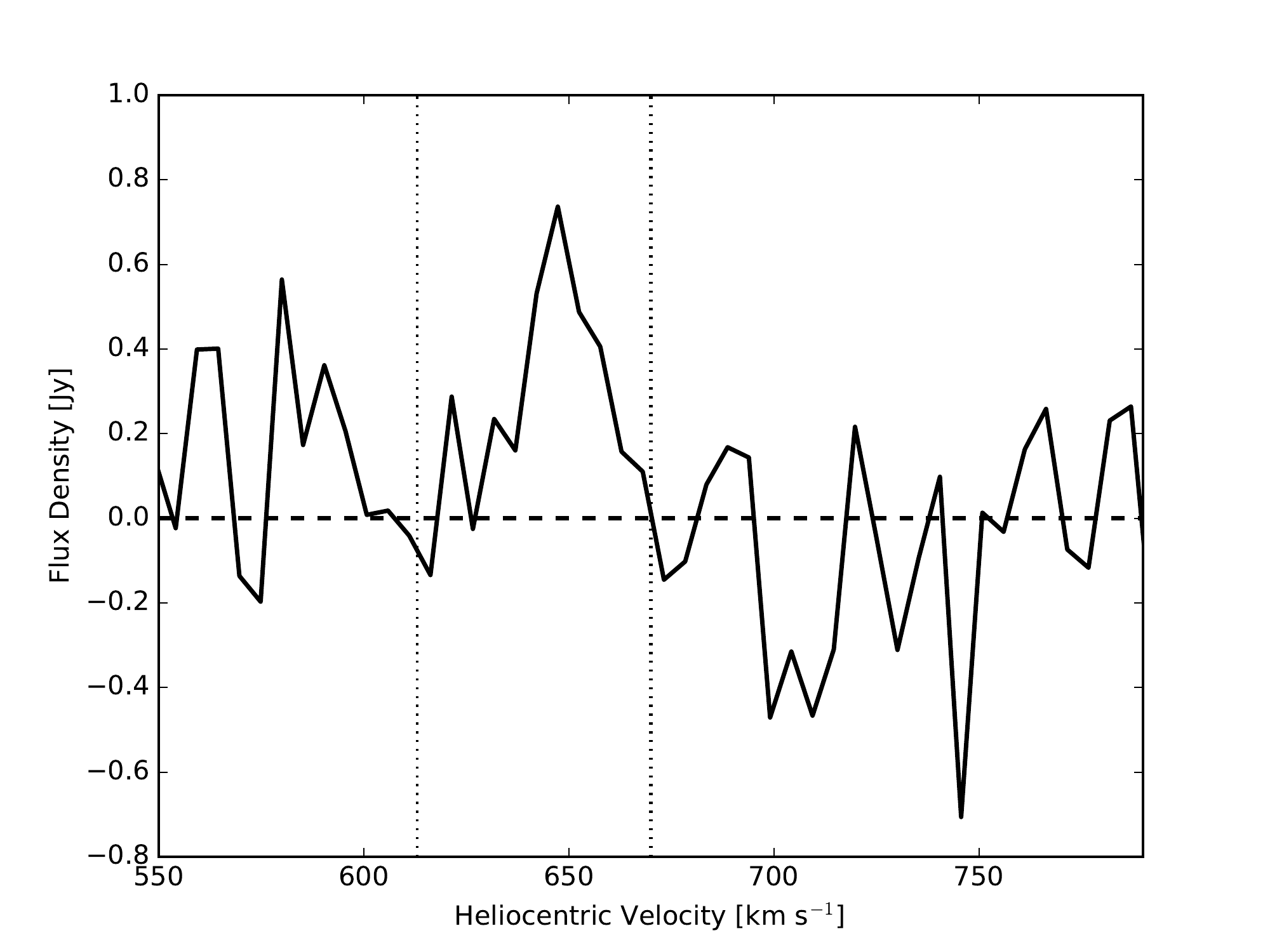}
\caption{Global $\hi$ profile derived from GBT data for the northern companion of NGC925 deemed `Halogas' by \citet{kara13}.}
\label{fig:NGC925CompProfile}
\end{figure}

Our GBT observations detect (5.79$\pm$0.29)$\times$10$^{9}$$\Msun$ of $\hi$, while the WSRT data reveal (5.54$\pm$0.28)$\times$10$^{9}$$\Msun$ of $\hi$ measured over the same angular area. H11 found faint emission extending towards the systemic velocity (see their Figure 6). This ``beard'' gas is interpreted as a slowly rotating halo seen in projection against the disk with a total $\hi$ mass on the order of 10$^{8}$ $\Msun$. While slight in magnitude, there is some emission picked up by the GBT which was missed in the original HALOGAS observations. 
The GBT observations therefore reveal additional $\hi$ in NGC925 that must be some combination of extended and diffuse as it was not detected in neither optimally weighted VLA data nor in the more sensitive WSRT data. We defer correcting the WSRT data for missing short spacings for a future paper; nevertheless, a high-resolution cube that recovers emission at angular scales will provide an excellent data set for a detailed dynamical study relating the beard gas to the extraplanar component.

H11 also noted the presence of the a small companion centered to the North at about $\alpha$$_{J2000}$=02$^{h}$27$^{m}$20$^{s}$, $\delta$=33$^{\circ}$57$'$30$''$ in the velocity range of 613-665 km s$^{-1}$. This companion was cataloged as `Halogas' by \citet{kara13}. The companion has a total $\hi$ mass (as measured by the GBT) of (3.11$\pm$0.15)$\times$10$^{7}$$\Msun$ consistent with the measurement from H11. A global $\hi$ profile taken from the GBT data is shown in Figure~\ref{fig:NGC925CompProfile}; there is no detection of a bridge of $\hi$ between Halogas and NGC925.

\subsection{NGC4414}\label{subsection:n4414}
NGC4414 is a moderately inclined ($i$ = 50$^{\circ}$) SAc galaxy, and is one of the most distant galaxies in the HALOGAS survey at 18$\pm$2 Mpc. NGC4414 was characterized through tilted ring fitting \citep{deBlok14} as having a regular rotating inner disk within 240$''$ ($\sim$21 kpc at the distance of NGC4414) in radius and an outer disk that is mostly dominated by rotation with some evidence for radial and noncircular motions towards the edge of the extended $\hi$ distribution. The high-resolution WSRT-HALOGAS observations do not show any evidence for an interaction besides the disturbed outer disk, though NGC4414 likely has undergone some weak interaction with neighboring galaxies within the Coma I cluster in the past. \citet{deBlok14} mention a possible source of an interaction may be the small galaxy, SDSS J122646.27+311904.8, but neither the WSRT or GBT detect any $\hi$ at its position. 

\begin{figure*}
\includegraphics[width=7 in]{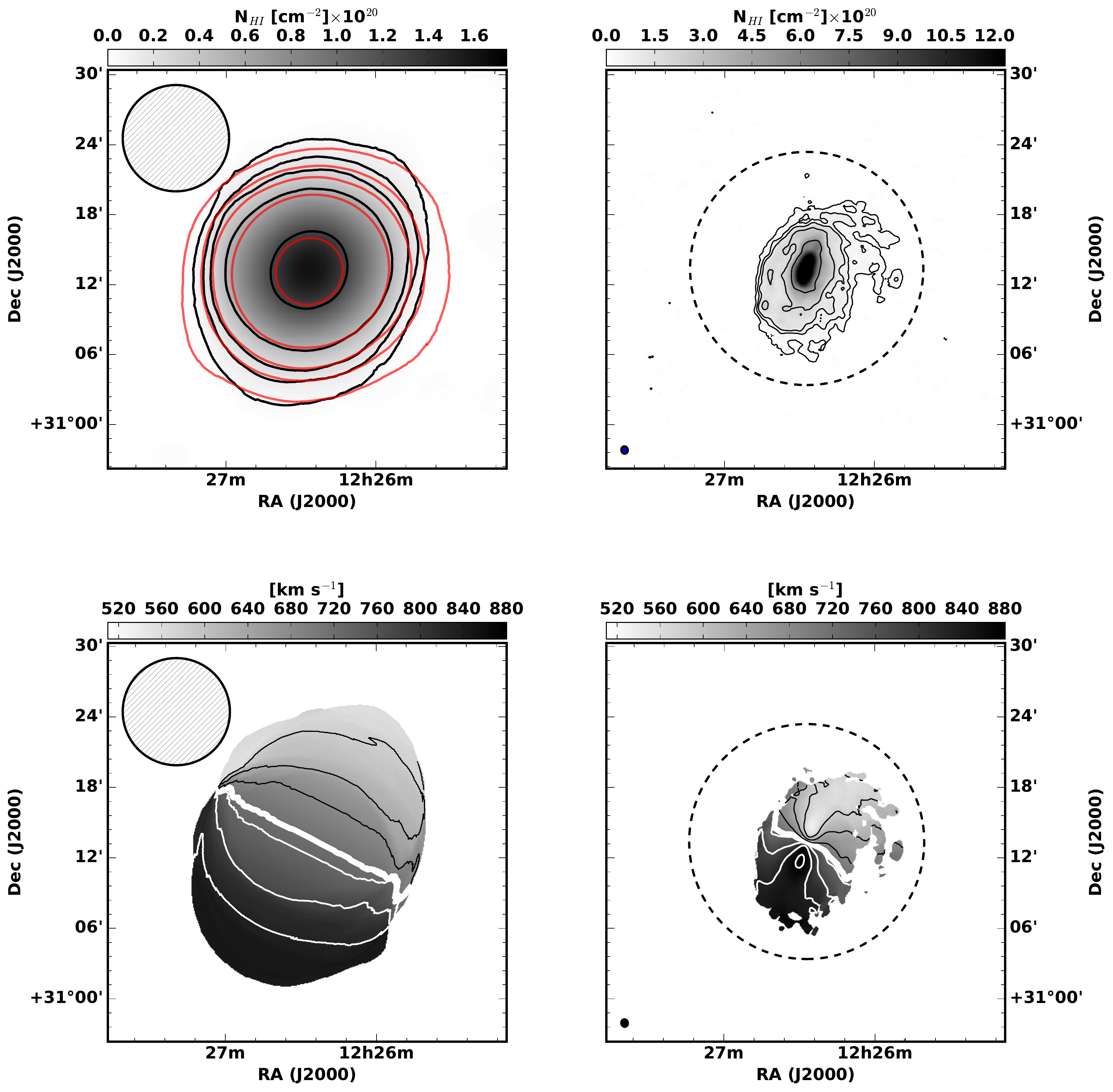}
\caption{A comparison of the $N_{HI}$ images (top) and velocity fields (bottom) of NGC4414 derived from the GBT (left) and WSRT-HALOGAS (right) data. The red and black contours in the low resolution $N_{HI}$ images respectively denote WSRT and GBT data; they begin at a $N_{HI}$ level of 5.0$\times$10$^{18}$ cm$^{-2}$ and increase by factors of 3, 5, 10, and 25. The contours in the high-resolution WSRT $N_{HI}$ image begin at a $N_{HI}$ level of 2.0$\times$10$^{19}$ cm$^{-2}$ and increase by factors of 3, 5, 10, and 25. The velocity contours in both maps begin at 530 km s$^{-1}$ and increase by 50 km s$^{-1}$ with the systemic velocity of NGC4414 of 716 km s$^{-1}$ represented by the thick white contour. The receding and approaching sides are denoted by white and black contours, respectively. The GBT (WSRT) beams are shown for their respective data sets in the top (bottom) left corner of each panel. The dashed circles in the right-hand panels represent the maximum recoverable angular scales of the WSRT data.}
\label{fig:n4414ContAll}
\end{figure*}

\begin{figure*}
\centering{\includegraphics[width=7.5 in]{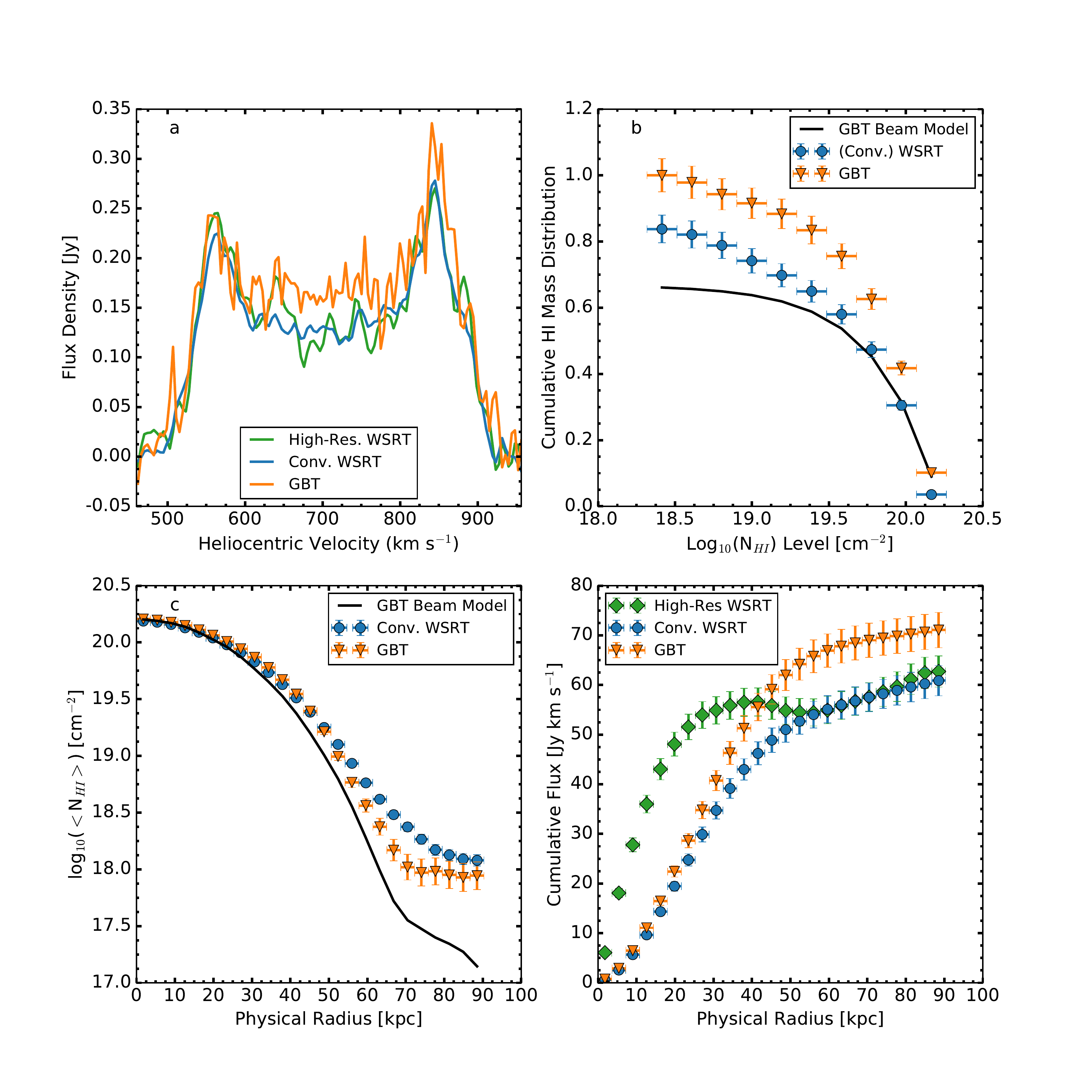}}
\caption{Comparison between the high-resolution (green diamonds) WSRT, convolved WSRT (blue circles), and  regridded GBT (orange inverted triangles) for NGC4414. $\textit{a}$: global $\hi$ profile; $\textit{b}$: Cumulative $\hi$ as a function of $N_{HI}$. The dashed and solid black lines simulate the contribution of a Gaussian beam and our GBT beam model, respectively; $\textit{c}$: projected physical radial dependence on the azimuthally averaged $N_{HI}$; $\textit{d}$: projected radial dependence of the cumulative flux. In this case, we also show the results of our analysis on the high-resolution WSRT data.}
\label{fig:n4414_4PanelSummary}
\end{figure*}

\begin{figure*}
\includegraphics[width=7 in]{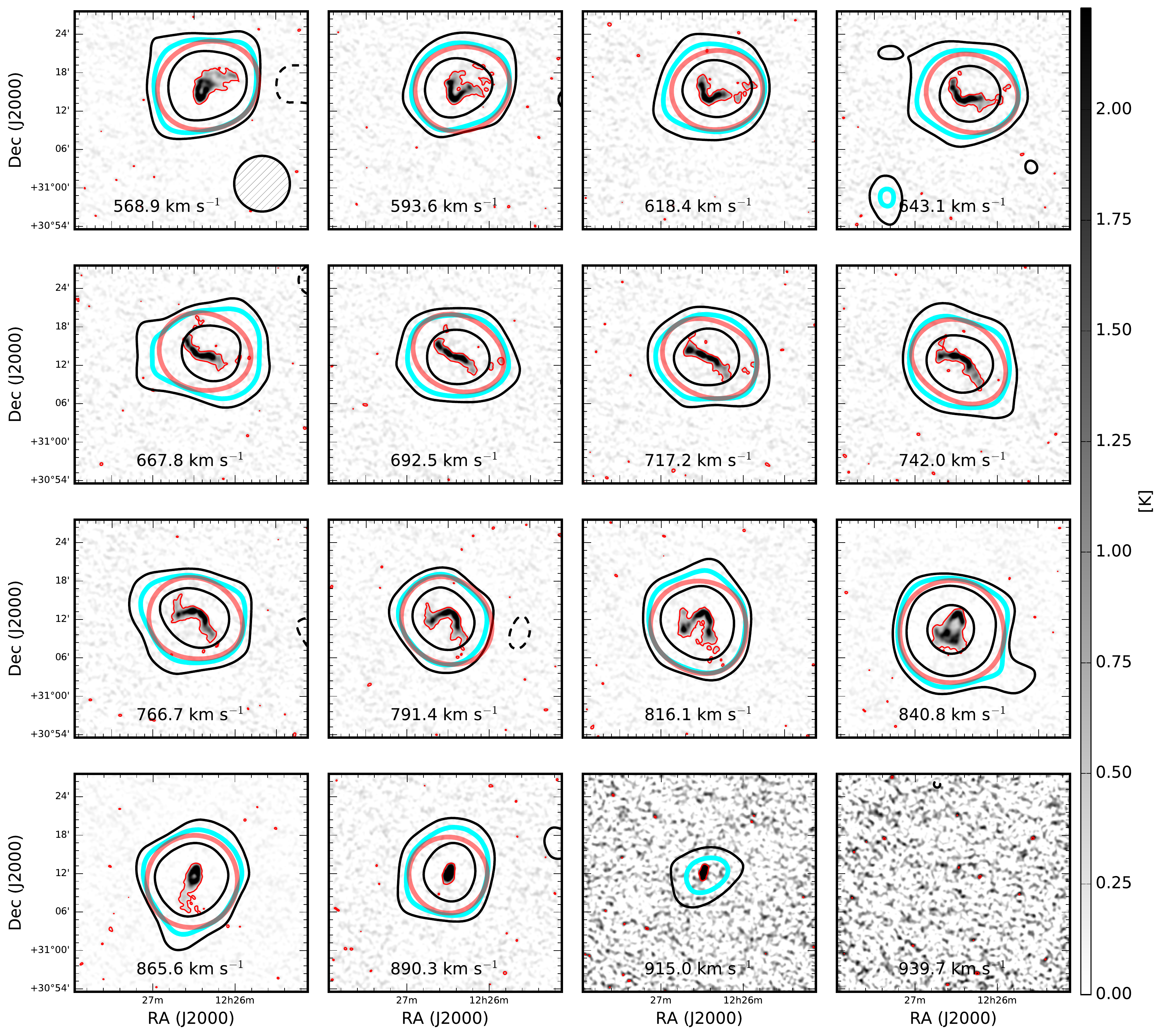}
\caption{Selected channel maps of the NGC4414 WSRT data cube with corresponding GBT channel maps superimposed. The GBT data are shown in black and cyan contours at levels of -3 (dashed), 3, 5 (thick cyan), and 25 times 0.01 K, or equivalently  a column density level of 9.75$\times$10$^{16}$ cm$^{-2}$ per 4.12 km s$^{-1}$ velocity channel. The grey-scale shows the $\hi$ emission from the WSRT cube. The thin red line  denotes a brightness temperature of 0.39 K, or a column density levels at 2.92$\times$10$^{18}$ cm$^{-2}$, and the thick red line denotes emission at 2.93$\times$10$^{17}$ cm$^{-2}$ (the same level of the cyan contour) in the primary-beam corrected WSRT cube convolved down to the GBT resolution. The GBT beam is shown in the top left panel.}
\label{fig:NGC4414ChanMaps}
\end{figure*}

A comparison between the GBT and WSRT data is shown in Figure~\ref{fig:n4414ContAll} in the form of integrated $\hi$ intensity and velocity field maps. As noted by \citet{deBlok14}, the inner regions of the velocity field show very well-behaved rotation in the inner regions of the galaxy with motions deviating from circular towards the edge of the $\hi$ disk. These areas of irregular rotation overlap well with the presence of the disturbed structure seen in the high resolution high-resolution WSRT $N_{HI}$ image, which again, may be evidence for a past interaction. The extent of the overall $\hi$ distribution is within the maximum recoverable angular scale as denoted by the dashed circles.

The global flux density profiles of NGC4414 derived from the GBT and WSRT data are shown in the top left panel of Figure~\ref{fig:n4414_4PanelSummary}. The GBT detects more flux overall, and excess flux is encountered over almost the entire velocity range. The total $\hi$ mass measured by the GBT is (5.43$\pm$0.27)$\times$10$^{9}$ $\Msun$, which is $\sim$1$\times$10$^{9}$ $\Msun$ more $\hi$ than is measured in the WSRT data over the same area. The large offset in the fluxes may be due, in part, to a large plume of $\hi$ extension to the West, which is further evidence for a past interaction.

The cumulative $\hi$ mass as a functions of $N_{HI}$ calculated from both data sets for NGC4414 are summarized in the top right panel. The convolved WSRT data begin to trace the simulated unresolved observation quite well past $log_{10}\left(N_{HI}/cm^{2}\right)$ = 19.0. The offset of $\sim$15\% between the two data profiles at the lower $N_{HI}$ bins indicates that the WSRT observations may resolve out structure that extends past the maximum recoverable angular scale with a peak $N_{HI}$ on the order of $\sim$10$^{18}$ cm$^{-2}$.

The radial dependent properties are shown in the lower two panels of Figure~\ref{fig:n4414_4PanelSummary}. The azimuthally averaged $N_{HI}$ profiles interestingly begin to differ at projected physical of about 60 kpc, while a prevalent dip in the cumulative flux profiles of the WSRT data begins near 45 kpc. Considering the $N_{HI}$ profiles are derived from masked $N_{HI}$ images, the deviation between the GBT and WSRT profiles could reflect the presence of artifacts, such as leftover residual sidelobes from deconvolution, that increase noise properties at larger angular extents to mimic the presence of legitimate signal. The variations in the WSRT cumulative flux profiles explicitly demonstrate the presence of artifacts in these data cubes.

Channel maps of NGC4414 similar to Figures~\ref{fig:NGC891FullChanMaps} and~\ref{fig:NGC925FullChanMaps} are presented in Figure~\ref{fig:NGC4414ChanMaps}. The outermost 3$\sigma$ contour generally extends past the 5$\sigma$ contour of the convolved WSRT data in each selected velocity channel. This is consistent with the behavior of the global $\hi$ profile from Figure~\ref{fig:n4414_4PanelSummary}. The channel maps reveal the extent of the NGC4414 emission in a single velocity channel is within the angular sensitivity limit ($\sim$20$'$, or $\sim$ 100 kpc at the distance of NGC4414) in the WSRT data. Additionally, the angular extent of NGC4414 is only a few times the GBT beam area as evidence by the steep drop in the cumulative $\hi$ mass function. Considering all these factors, we conclude that a majority of the excess emission detected by the GBT is likely due to artifacts (e.g., residual sidelobe structure) in the WSRT cubes, as opposed to resolved out structure. 

\citet{deBlok14} noted in their analysis that gas at velocities lower than the local rotational velocity (i.e., "beard" gas) is present in both the receding and approaching sides of the rotation curve for the inner disk of NGC4414, and should be interpreted as extraplanar gas in the inner disk. Through a variety of techniques including blanking high column density pixels and subtracting Gaussian fits of the velocity profiles within the observed data cube \citep{frat02}, \citet{deBlok14} determined the $\hi$ mass of the extraplanar gas associated with the inner disk to be between 2 and 6.5 percent of the total $\hi$ mass of the inner disk (i.e., about 2$\times$10$^{8}$$\Msun$). Similarly, the authors concluded the extraplanar gas associated with the outer disk to be about 1 to 2 percent of the total $\hi$ mass of the system. In the future, merging the GBT and WSRT data sets will provide an excellent basis for characterizing dynamical links between anomalous velocity gas and an extraplanar component in NGC4414 as well as characterizing the kinematics of a potential past interaction.

\subsection{NGC4565}\label{subsection:n4565}

\begin{figure*}
\includegraphics[width=7 in]{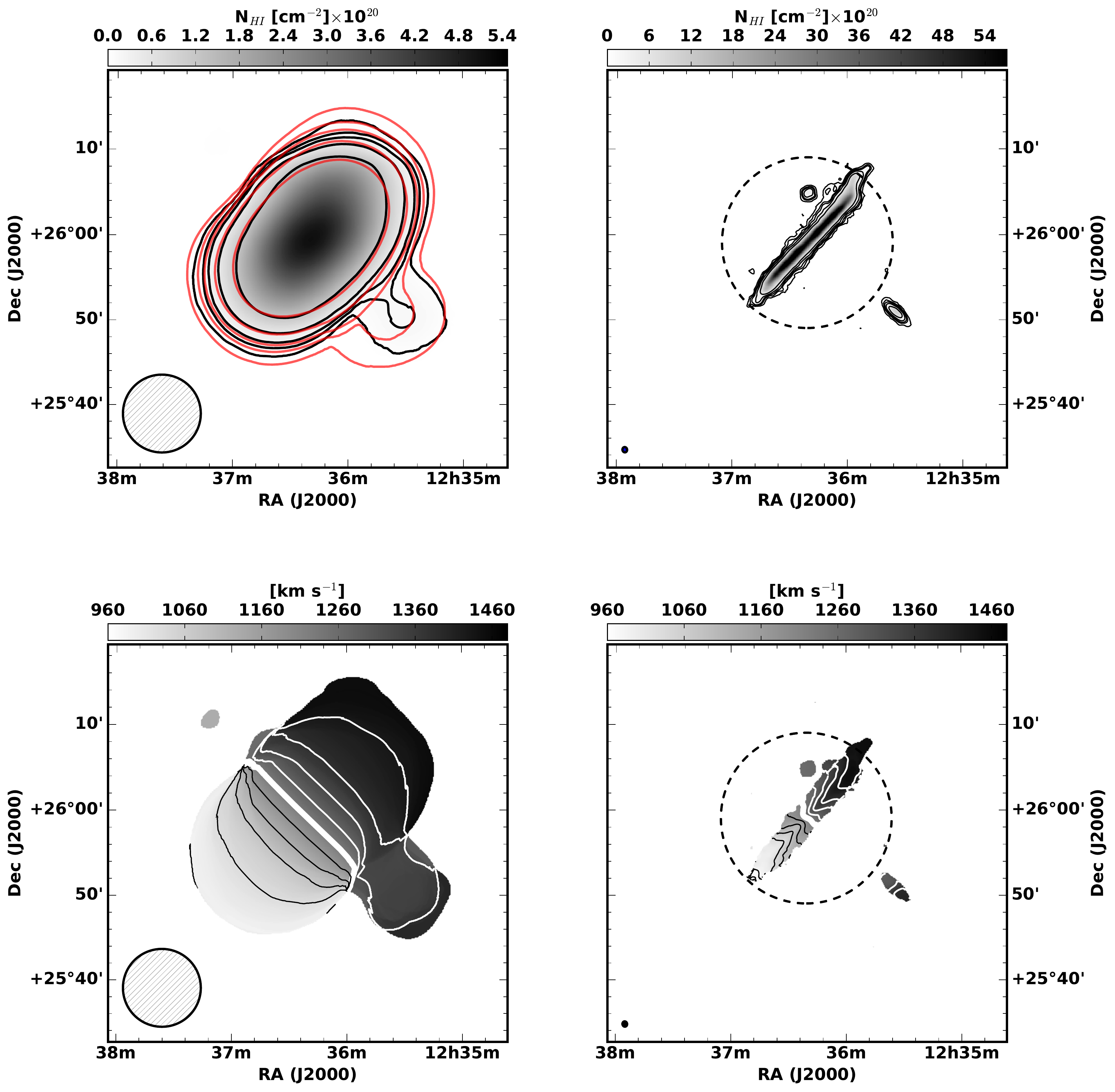}
\caption{$N_{HI}$ images (top) and velocity fields (bottom) for GBT (left) and WSRT (right) data for NGC4565. The red and black contours in the low resolution $N_{HI}$ images respectively denote WSRT and GBT data. The contours start at a column density value of 5$\times$10$^{18}$ and continue at 3, 5, 10, and, 25 times that level in the GBT image. The contours in the associated high-resolution WSRT image begin at a level equivalent to 2.0$\times$10$^{19}$ cm$^{-2}$ and continue at 3, 5, 10, and 25 times that level.  The contours in both velocity fields begin at 970 km s$^{-1}$ and continue in steps of 50 km s$^{-1}$. The systemic velocity of 1230 km s$^{-1}$ is represented by the thick line, and the approaching and receding velocities are denoted by black and white contours, respectively. The dashed circles in the right-hand panels represent the maximum recoverable angular scales of the WSRT data.}
\label{fig:n4565ContAll}
\end{figure*}

\begin{figure*}
\centering{\includegraphics[width=7.5 in]{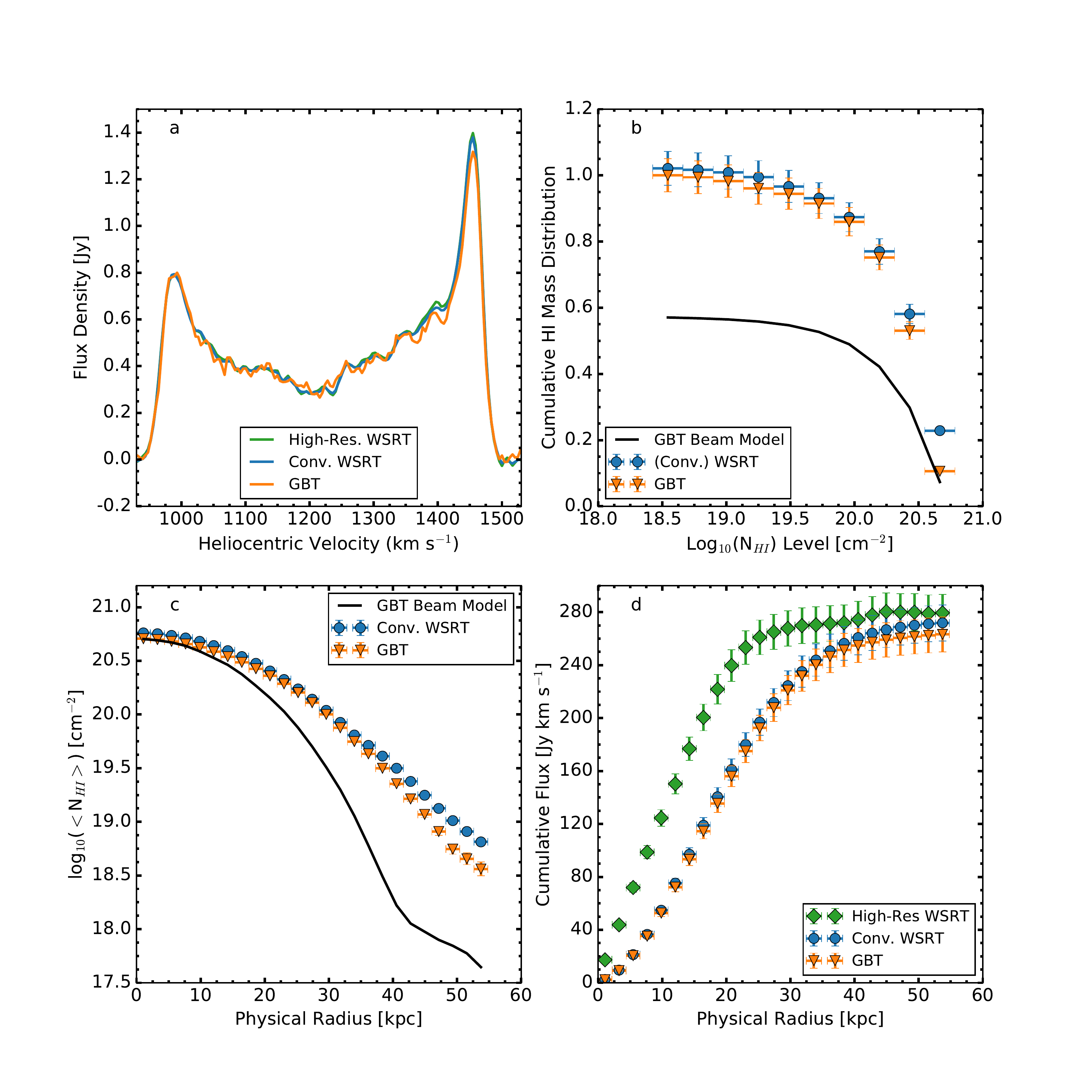}}
\caption{Comparison between the high-resolution (green diamonds) WSRT, convolved WSRT (blue circles), and  regridded GBT (orange inverted triangles) for NGC4565. $\textit{a}$: global $\hi$ profile; $\textit{b}$: Cumulative $\hi$ mass as a function of $N_{HI}$. The dashed and solid black lines simulate the contribution of a Gaussian beam and our GBT beam model, respectively; $\textit{c}$: projected physical radial dependence on the azimuthally averaged $N_{HI}$; $\textit{d}$: projected radial dependence of the cumulative flux. In this case, we also show the results of our analysis on the high-resolution WSRT data.}
\label{fig:n4565_4PanelSummary}
\end{figure*}

\begin{figure*}
\includegraphics[width=7 in]{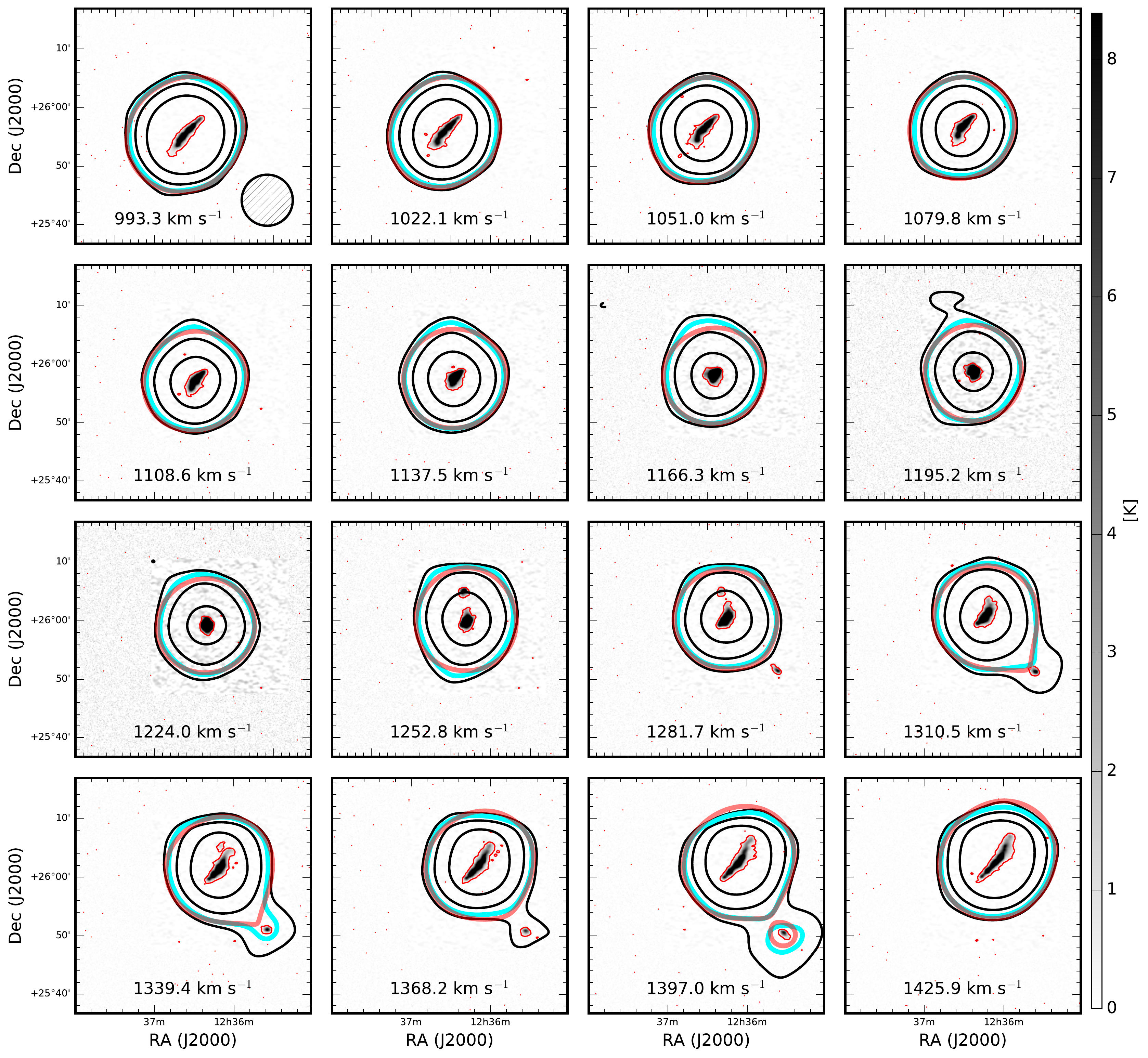}
\caption{Selected channel maps of the NGC4565 WSRT data cube with corresponding GBT channel maps superimposed. The GBT data are shown in black and cyan contours at levels of -3 (dashed), 3, 5 (thick cyan), and 25 times 0.01 K, or equivalently  a column density level of 1.12$\times$10$^{17}$ cm$^{-2}$ per 4.12 km s$^{-1}$ velocity channel. The grey-scale shows the $\hi$ emission from the WSRT cube. The thin red line  denotes a brightness temperature of 0.54 K, or a column density levels at 4.05$\times$10$^{18}$ cm$^{-2}$, and the thick red line denotes emission at 5.60$\times$10$^{17}$ cm$^{-2}$ (the same level of the cyan contour) in the primary-beam corrected WSRT cube convolved down to the GBT resolution. The GBT beam is shown in the top left panel.}
\label{fig:NGC4565FullChanMaps}
\end{figure*}

NGC4565 is a large, edge-on SAb galaxy with two very nearby companion galaxies: IC 3571 directly to the north and NGC 4562 to the southwest; only NGC 4562 is partially resolved in our GBT data. The apparent connection is due to the beam confusion stemming from the large GBT beam. The SFR for NGC4565 is among the lowest in the entire HALOGAS sample at 0.67 $\Msun$ yr$^{-1}$. The $N_{HI}$ images and velocity fields derived from both the GBT and WSRT data sets are detailed in Figure~\ref{fig:n4565ContAll}. These specific HALOGAS observations for NGC4565 have been discussed by \citet{zsch12} and expose a potential interaction between IC3571 and NGC4565 as evidenced by the possible tidal material shown as a separate cloud complex between IC3571 and the disk of NGC4565 in the top right panel of Figure~\ref{fig:n4565ContAll}. It remains unclear if this structure is related to tidal interactions between the main disk of NGC4565 and IC3571, or an accretion process.

The total $\hi$ measured in the GBT data over the same angular region as the WSRT data (including companions) is (7.46$\pm$0.39$)\times$10$^{9}$$\Msun$, while the primary-beam corrected WSRT data convolved to the GBT resolution gives (7.32$\pm$0.37$)\times$10$^{9}$$\Msun$. The $\hi$ profiles are shown in Figure~\ref{fig:n4565_4PanelSummary} and match extremely well.

The cumulative $\hi$ mass as a function of $N_{HI}$ for NGC4565 is presented in Figure~\ref{fig:n4565_4PanelSummary}b. Both distributions also trace each other remarkably well within statistical uncertainties. Additionally, the excellent agreement between all NGC4565 data sets extend to the mean radial column density/cumulative flux. The slight offsets towards higher projected physical radii can be attributed to a residual baseline structure in the GBT cube.

Selected channels are shown in Figure~\ref{fig:NGC4565FullChanMaps} with the GBT and WSRT superimposed on the high-resolution WSRT channels. The 5$\sigma$ level of emission detected from the GBT is consistent with the same level in the convolved WSRT data. 

The consistency in the various profiles of Figure~\ref{fig:n4565_4PanelSummary} and contours in Figure~\ref{fig:NGC4565FullChanMaps} indicate there is no extended diffuse $\hi$ reservoir around NGC4565 at the $N_{HI}$ levels of $\sim$10$^{18}$ cm$^{-2}$. \citet{zsch12} also did not find evidence using the same high resolution WSRT data of any significant amount of extraplanar $\hi$. The authors of that study proposed this absence is due to no significant disk-halo cycling of material in NGC4565. The lack of any sort of extended low column density $\hi$ component supports this conclusion.

\section{Discussion}\label{section:Discussion}
In this section we will first discuss our initial survey results in the context of future $\hi$ surveys planned for forthcoming radio telescopes. The second subsection explores methods to link the diffuse environment of the HALOGAS galaxies to intrinsic galaxy properties and thus the theoretical predictions from simulations. 

\subsection{Implications for Future $\hi$ Surveys}\label{subsection:implications}

It is well known through a combination of models (e.g., \citealt{mal93,dove94, blandHawth17}) and observations \citep{vanGor93} that the transition from optically-thick to optically-thin medium decreases the efficiency of self-shielding leading to a dramatic drop in the fraction of $\hi$ to total hydrogen (i.e., neutral fraction) primarily due to UV and X-Ray background radiation. This effect is demonstrated in cosmological simulations of the $\hi$ distribution (e.g., \citealt{pop09, Rahmati15, Marinacci17}); in particular, \citet{pop09} shows the neutral fraction drops from unity to about a percent between 18.0 $<$ $log_{10}$($N_{HI}$/cm$^{2}$) $<$ 20.0. This rapid decrease results in a plateau in the $\hi$ distribution function that effectively predicts the probability of detecting substantial $\hi$ structures at $N_{HI}$$\sim$10$^{18}$ cm$^{-2}$ is very low even with deep surveys. The flux as a function of velocity, cumulative $\hi$ mass as a function of $N_{HI}$, and azimuthally averaged profiles of column density and cumulative flux show that, overall, the WSRT observations do an excellent job of recovering the low level $\hi$. Save for NGC4414, we see that the $\hi$ distributions do not significantly differ between $N_{HI}$$\sim$10$^{18}$ cm$^{-2}$ and $\sim$10$^{19}$ cm$^{-2}$.

The dearth of diffuse $\hi$ around these sources has intriguing implications on both the search for cold flow accretion and the determination of the edge of disk gas in late-type galaxies. Many simulations (e.g., \citealt{keres05,keres09, birnDek03}) have predicted that less massive galaxies at low redshift will replenish gas through the form of cold filaments which penetrate the halo down to the disk edge where some fraction of the gas should cool sufficiently to be neutral. Aside from characterizing gas accretion mechanisms, deep $\hi$ observations are also useful for tracing large-scale structure in the local Universe. Simulations undertaken by \citet{nuza14} predict M31 and Milky Way analogues to sit within a clumpy, yet very diffuse neutral component with $\hi$ masses on the order of $\sim$10$^{8}$ $\Msun$ existing predominately above $N_{HI}\sim 10^{17}$ cm$^{-2}$. Our cumulative $\hi$ mass as a function of $N_{HI}$ plots are consistent with this prediction. On average, only 2\% of the total $\hi$ mass exists in column densities below 1$\times$10$^{19}$ cm$^{-2}$ as measured in the low-resolution data sets (see Section~\ref{subsection:cmaRelation}). Given the results of~\citet{pop09}, in which the $\hi$ surface area distribution function is predicted to steepen at $N_{HI}$ levels lower than 10$^{18}$ cm$^{-2}$, we expect the offset between the GBT and WSRT cumulative $\hi$ distributions would widen considerably with the detection of diffuse $\hi$ that traces the large scale structure of the cosmic web, an extended reservoir of diffuse neutral $\hi$, and/or cold flow filaments at the 10$^{18}$ cm$^{-2}$ column density level. Such structures would likely be resolved out in an interferometric data set and therefore not contribute at all to the cumulative $\hi$ mass at these low $N_{HI}$ levels.
 
Detailed analysis of the high-resolution HALOGAS $\hi$ data by E. J\"utte (2018; in preparation) reveal only a few newly detected $\hi$ clouds which may be accreting. Furthermore, maps produced by \citet{wolfe16} of the apparent connection between M31 and M33 original discovered by \citet{brth04}, which go several times deeper in column density than our observations, reveal only discrete clouds of $\hi$ with masses on the order of $\sim$10$^{5}$ $\Msun$ as opposed to a continuous, smooth diffuse component. These early high-resolution results coupled with the small amount of newly detected $\hi$ in the initial GBT data suggest that, if these galaxies are currently accreting gas, most of it must be ionized and unobservable in $\hi$ emission. The absence of significant $\hi$ structure at the $N_{HI}$ level of 10$^{18}$ cm$^{-2}$ supports the notion that $\hi$ surveys of external galaxies must go $\textit{at least}$ as deep as $N_{HI}$$\sim$10$^{17}$ cm$^{-2}$ to substantially increase the probability of detecting emission associated with the IGM and/or cold mode accretion. 
 
\subsection{The Relationship Between Diffuse $\hi$ and Cold Mode Accretion}
\label{subsection:cmaRelation}
\begin{figure*}
\centering{
\includegraphics[width=7in] {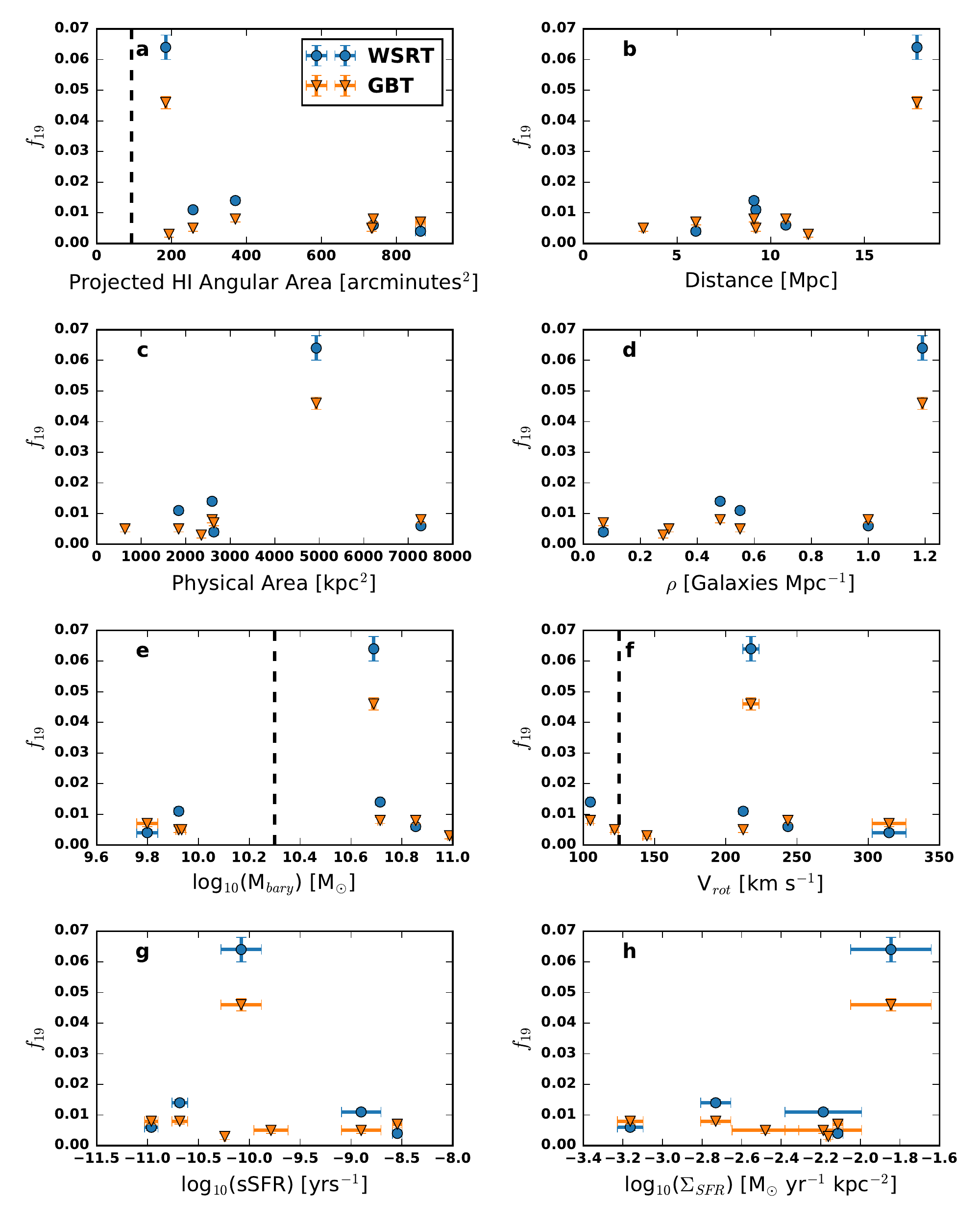}
\caption{Environmental and intrinsic property relations with excess $\hi$ ratio, $f_{19}$. a: $f_{19}$ as a function of projected angular area. The vertical dashed line marks the area of the GBT beam (up to the FWHM); b: $f_{19}$ as a function of distance; c: $f_{19}$ as a function of projected physical area; d: $f_{19}$ as a function of galaxy density. $\rho$ is taken from \citet{tully88}; e: $f_{19}$ as a function of the source galaxy baryonic mass (M$_{HI}$ + M$_*$). The vertical dashed line corresponds to the baryonic mass threshold from \citet{keres05} below which cold mode accretion is the dominant accretion mechanism; f: $f_{19}$ as a function of rotational velocity. The vertical dashed line corresponds to the threshold mass at which galaxies should be gas dominated \citep{kann13}; g: $f_{19}$ as a function of specific SFR (sSFR); h: $f_{19}$ as a function of the surface density of star formation ($\Sigma_{SFR}$).}
\label{fig:environ}}
\end{figure*}

While the GBT data do not reveal new features below the $N_{HI}$$\sim$10$^{18}$ cm$^{-2}$ level, as the remaining data are analyzed it is still important to consider the implications of potential diffuse features and their relation to cold flows as predicted from simulations. Namely, how would diffuse $\hi$ related to cold mode accretion manifest itself in emission? Does the diffuse gas originate from outflows, tidal interactions with nearby companions, or accretion from the IGM? 

To explore the relationship between the diffuse $\hi$ around nearby galaxies and simulation predictions, we measure the diffuse $\hi$ mass fraction as defined by 
\begin{equation}
f_{19} = 1 - \frac{M_{19}}{M_{HI}}, 
\end{equation}
where $M_{HI}$ is the total $\hi$ mass detected for that particular telescope (taken from pixels with $N_{HI}$ values at and above the 3$\sigma$ detection limit listed in Table~\ref{tab:mapPropSummary} to avoid measuring into the noise) and $M_{19}$ is the $\hi$ mass above the $N_{HI}$ level of 1$\times$10$^{19}$ cm$^{-2}$ in the unmasked GBT and convolved WSRT $N_{HI}$ images. 

The purpose of this parameter is to measure the mass fraction of diffuse $\hi$ associated with a given galaxy. We can then relate the presence (or dearth) of diffuse $\hi$ to properties predicted by cosmological simulations to correlate with higher rates of cold mode accretion. In the context of our subsequent discussion, we define diffuse $\hi$ to firstly be below the $N_{HI}$ level of 1$\times$10$^{19}$ cm$^{-2}$, which is on the same order as the analytical critical column densities derived by \citet{blandHawth17}, \cite{mal93}, and \citet{dove94} (i.e., where the $\hi$ transitions from mostly neutral to mostly ionized) and is approximately the 5$\sigma$ $N_{HI}$ detection limit of the native resolution WSRT data over a 20 km s$^{-1}$ line. Secondly, diffuse $\hi$ is implied to be extended over angular scales larger than the maximum recoverable angular scale of the WSRT ($\sim$20$'$). A lower $f_{19}$ value measured for a well-resolved source (relative to the larger GBT beam) would therefore demonstrate that most of the $\hi$ resides at higher column densities, indicating no signs of a diffuse inflowing component that can be observed in emission. On the other hand, a higher $f_{19}$ value would suggest the WSRT resolves out large scale $\hi$ features that possess $N_{HI}$ levels lower than $\sim$ 10$^{19}$ cm$^{-2}$.

Because $N_{HI}$ is directly proportional to the measured brightness temperature, which is itself averaged over the beam solid angle, the usefulness of the $f_{19}$ parameter to characterize the diffuse $\hi$ mass fraction relies on the sources being well-resolved. Once the source becomes unresolved, our measurement --- that is effectively the convolution of the true $\hi$ emission distribution with the larger beam --- will spread the flux of the out to angular scales where the beam response is lower, thus measuring low column density values that are no longer physical but still nevertheless begin to dominate the overall $\hi$ mass fraction and bias our quantity. While $f_{19}$ provides a quantity that can be related to properties predicted to be influential to accretion, given that the sources presented in this work are at different distances, span a range of physical scales, and are in general inhomogeneous, acute attention to several resolution indicators, which will be discussed below, is required to ensure a $f_{19}$ value for a given galaxy is not biased by resolution effects.

The $f_{19}$ values are summarized in Table~\ref{tab:derivedParams}. In general, the $f_{19}$ values derived from the convolved WSRT data are similar to their regridded GBT counterparts. In the cases where the convolved WSRT values are higher, the increase can be attributed to the increased column density sensitivity when the WSRT data (where emission fills the synthesized beam) is convolved to low resolution. In addition to the four sources discussed above, we also compute $f_{19}$ for three galaxies which have similar GBT data: NGC2403 (also a HALOGAS source) (\citealt{frat01, frat02, deBlok14}), NGC2997 \citep{pisano14}, and NGC6946 (with additional deep WSRT data; \citealt{pisano14, booms08}).

Our initial small sample size inhibits us from making any significant conclusions about the relations between fraction of diffuse $\hi$ and any of the properties which trace simulation predictions. However, investigating the trends within our initial small sample will provide intuition on any underlying correlations that may be revealed as we build up our sample and flesh out issues due to resolution effects.

As mentioned above, it is critical to rule out observational or resolution bias as a source for potential trends. Figure~\ref{fig:environ}a plots $f_{19}$ for each source as a function of projected $\hi$ angular area computed from the number of pixels above the associated 2$\sigma$ $N_{HI}$ level from Table~\ref{tab:mapPropSummary} in the unmasked high-resolution WSRT $N_{HI}$ images. Note that in each panel the blue circles denote $f_{19}$ values derived from convolved WSRT data and the orange inverted triangles represent GBT $f_{19}$ values. The vertical dashed line represents the area of the GBT beam above which extended source structure becomes resolved. While it is encouraging to see all of the projected HI angular area data points fall to the right of the GBT beam area, this is only one indication that our $f_{19}$ parameter is not biased by resolution effects. Because $f_{19}$ is essentially one minus the ratio of the integral of the cumulative $\hi$ mass distribution taken from bins larger than $Llg_{10}\left(N_{HI}/cm^{2}\right)$ = 19.0 to an integral taken over the full range of $N_{HI}$ bins, a better indicator for possible resolution bias is to ensure the data points of the cumulative $\hi$ mass distribution for a given source do not trace the simulated unresolved observation (i.e., the GBT beam profile).

In the case of Figure~\ref{fig:environ}a, the highest $f_{19}$ value is indeed associated with NGC4414, whose angular area is only a few times that of the GBT beam. Figure~\ref{fig:n4414_4PanelSummary}b shows the WRST data cumulative $\hi$ mass fraction is consistent with the observation of an unresolved source within the large GBT beam at higher $N_{HI}$ bins, while the GBT cumulative $\hi$ mass distribution shows only marginal improvement over the simulated unresolved observation. The slight trend between increasing distance and increasing $f_{19}$ values in Figure~\ref{fig:environ}b does hint that $f_{19}$ will be influenced by the large GBT beam. The high $f_{19}$ values in both data sets for NGC4414 reveal how the effects of resolution can bias the calculation of $f_{19}$. That said, we also see sources with comparable $f_{19}$ values over a large range of projected angular areas whose cumulative $\hi$ mass functions do not trace a simulated observation of an unresolved source. Figures~\ref{fig:environ}a and b therefore demonstrate that $f_{19}$ is not generally biased when applied to well-resolved sources.

The physical extent of a diffuse gas reservoir should ideally scale with a diffuse $\hi$ mass fraction. Plotting $f_{19}$ as a function of physical projected $\hi$ area in Figure~\ref{fig:environ}c shows that the highest $f_{19}$ value is associated with the second largest physical extent; however, the high $f_{19}$ value is very likely biased due to the greater distance of the source. On the other hand, the second highest $f_{19}$ value corresponds to the very well-resolved galaxy of NGC4565, demonstrating the expected scaling. That said, the fact this particular GBT $f_{19}$ value is the second highest value of the sample by only about a factor of two underscores the lack of a significant diffuse $\hi$ around NGC4565.

A relationship between $f_{19}$ and galaxy properties investigated in simulations (e.g., \citealt{keres05,keres09}) should give some indication as to whether any newly probed diffuse component could be related to accretion from the IGM. For example, simulations show cold mode accretion is still an available channel for gas accretion around nearby galaxies with relatively low halo masses ($M_{halo}$ $\leq$ 10$^{11.4}$ $\Msun$) and in low density environments. Any trend between excess $\hi$ and properties known to be responsible for the presence of excess $\hi$ could provide observational evidence, albeit indirect, that the observed excess $\hi$ is being accreted directly from the IGM. Figure~\ref{fig:environ}d plots $f_{19}$ as a function of galaxy density, or the number of galaxies per Mpc$^{-3}$ ($\rho$); we estimate $\rho$ by using values derived in \citet{tully88}. Any correlation between $f_{19}$ and $\rho$ should provide insight into the influence of tidal forces. No obvious trend presents itself here. The absence of a trend in the context of the full sample would be indicative of the relative unimportance of tidal interactions as compared to outflows or accretion to the presence of extraplanar and/or an extended gas reservoir. 

Figure~\ref{fig:environ}e plots $f_{19}$ as a function of the baryonic mass ($M_{bary}$) where the vertical dashed lines represent the threshold set by the simulations of \citet{keres05} where cold mode accretion is predicted to be the dominant mechanism of gas accretion. Assuming the dust and molecular gas components to be negligible to the total gas mass, we calculate $M_{bary}$ to be equal to $M_{HI}\cdot1.36 + M_{*}$, where $M_{HI}$ is measured by the GBT. The factor of 1.36 in the leading term corrects for the fraction of neutral He. There is no apparent trend between $f_{19}$ and $M_{bary}$. Additionally, Figure~\ref{fig:environ}f presents the relationships between $f_{19}$ and rotational velocity ($V_{rot}$). The dashed line in this plot shows the observational threshold such that galaxies with $V_{rot}$ $\leq$ 125 km s$^{-1}$ are expected to be gas dominated due to continuous sustained growth \citep{kann04,kann13}. Again, no discernible trend is observed.  

The final origin scenario for a large diffuse component in these galaxies is related to outflows due to star formation activity in the disk. The relationship between $f_{19}$ and specific star formation rate (sSFR), which is the star formation rate per unit stellar mass, is presented in Figure~\ref{fig:environ}g; $f_{19}$ as a function of the surface density of star formation is shown in Figure~\ref{fig:environ}h. We derive the surface density of star formation by dividing the SFR by the de-projected physical area. We compute the physical area assuming each galaxy face-on is roughly circular and utilize the de-projected and extinction corrected semimajor angular diameter computed from the B-band 25 mag arcsec$^{-2}$ isophote \citep{devac76}. Excluding the rightmost data points in Panel h (which may, at some level, be biased by resolution), there is no indication of trend between increasing $f_{19}$ with increased star formation activity. A positive correlation suggests more energy per gravitational potential in such galaxies, which points to the possible existence of outflows. 

Some, if not most, of the extraplanar gas observed around nearby galaxies must originate from outflows described by galactic fountain (\citealt{breg80,norm89, frat17}), and are driven by the injection of momentum and energy into the local interstellar medium from supernovae. In most cases, such outflows are the more feasible origin for diffuse $\hi$ in the form of extraplanar gas. \citet{frat06} and \citet{frat08} outline dynamical models that accurately reproduce the general vertical distribution of gas in NGC891 and NGC2403 with a necessary energy injection of only $<$4\% of the energy released by supernovae. But while the general distribution is reproduced, the rotational velocity of the model halos is much too high compared to observations. The discrepancy between modeled and observed rotational velocities suggested the acquisition of low-angular momentum gas from the IGM in order to slow the model halos. More recently, however, \citet{marinacci10} and \citet{frat17} show that including the effects of condensation, drag (ram pressure), and radiative cooling in these ballistic models of fountain gas will prevent momentum loss in the cold gas while also increasing the transfer of momentum to intermediate temperature material as opposed to the hot coronal gas. The increase in rotational velocity is therefore mitigated, thus bringing theses models into agreement with observational data. By incorporating radiative cooling of the hot coronal gas, it can condense in the wake of fountain clouds which consequently raises the accretion rate of low-metallicity gas onto the central disk of an embedded galaxy. Specifically, \citet{frat17} recover cold gas accretion rates comparable with the the SFR of NGC891 and NGC2403. 

We end this section by posturing what type of observational constraints between $f_{19}$ and the inherent galaxy properties shown in Figure~\ref{fig:environ} would indicate the possibility of observing active cold mode accretion. The first necessity would be a $f_{19}$ value from the GBT data that is well above the WSRT and well to the right of the vertical dashed line in Figure~\ref{fig:environ}a. This would indicate that the WSRT had resolved out significantly extended structure and physical column densities as the emission has filled the large GBT beam. That same large $f_{19}$ GBT data point would then need to be associated with a small $\rho$ to indicate isolation and that tidal interactions are not likely to be the origin of the diffuse environment. We would next need to see the $f_{19}$ value fall below the $M_{bary}$ and $V_{rot}$ thresholds to ensure the galaxy is consistent with the predictions of the cosmological simulations. Finally, the presence of a significant diffuse $\hi$ component would need to be associated with a relatively insignificant amount of star formation activity ruling out the presence of substantial outflows and subsequent condensation from the fountain activity.

The derived properties of these four sources are summarized in Table~\ref{tab:derivedParams}. The columns are (1) Source Name; (2) $\hi$ mass measured in the GBT data; (3) $\hi$ mass measured in the convolved WSRT data; (4) $\hi$ mass as measured in the high-resolution WSRT data; (5) stellar mass derived from the $WISE$ All Sky Image Atlas; (6) baryonic mass; (7) deprojected physical area; (8) rotation velocity from HYPERLEDA (9) $f_{19}$ measured in the convoled WSRT cube; and (10) $f_{19}$ measured in the GBT cube. The relationships in this section should be considered diagnostic in that we are trying to develop intuition into what trends should exist in galaxies that are actively undergoing cold mode accretion. The low number of sources studied in this work inhibit any conclusions. We will continue to build our statistical sample with future analysis of the HALOGAS, THINGS, and ultimately the \textit{MeerKAT HI Observations of Nearby Galactic Objects: Observing Southern Emitters} (MHONGOOSE; \citealt{deBlok17}) and the \textit{Imaging Galaxies Inter-galactic and Nearby Environment} (IMAGINE) galaxy samples in order to search for correlations between the presence of a significant diffuse $\hi$ environment and the intrinsic galaxy properties which should trace accretion from the IGM in the local Universe.

\section{Conclusions and Future Work}\label{section:conclusion}

\begin{table*}
	\centering
    \resizebox{\textwidth}{!}
{\begin{tabular}{lccccccc}
    \hline \hline
	Derived Properties & NGC891 & NGC925 & NGC2403 & NGC2997 & NGC4414 & NGC4565 & NGC6946 \Tstrut\Bstrut\\
    \hline
    GBT $\hi$ Mass [10$^{9}$$\Msun$] & 3.86$\pm$0.19 & 5.79$\pm$0.29 & 3.39$\pm$0.37 & 7.0$\pm$1.0 & 5.43$\pm$0.27 & 7.33$\pm$0.37 & 3.80$\pm$0.69 \Tstrut\Bstrut\\
    Conv. WSRT $\hi$ Mass [10$^{9}$$\Msun$] & 3.81$\pm$0.19 & 5.54$\pm$0.28 & --- & --- & 4.56$\pm$0.22 & 7.46$\pm$0.39 & --- \Tstrut\Bstrut\\
    High-Res. WSRT $\hi$ Mass [10$^{9}$$\Msun$] & 3.90$\pm$0.18 & 5.57$\pm$0.28 & --- &  --- & 4.63$\pm$0.23 & 7.55$\pm$0.38 & ---  \Tstrut\Bstrut\\
    M$_{*}$ [10$^{10}$$\Msun$] & 0.30$\pm$0.10 & 4.40$\pm$0.10 & 0.39$\pm$0.01 & 8.70$\pm$0.10 & 4.15$\pm$0.10 & 6.15$\pm$0.10 & 0.12$\pm$0.01 \Tstrut\Bstrut\\
    M$_{bary}$ [10$^{10}$$\Msun$] & 0.82$\pm$0.02 & 5.20$\pm$0.10 & 0.86$\pm$0.03 & 9.71$\pm$0.18 & 4.89$\pm$0.10 & 7.15$\pm$0.10 &  6.29$\pm$0.59  \Tstrut\Bstrut\\
    Deprojected Physical Area [kpc$^{2}$] & 603$\pm$6 & 490$\pm$6 & 122$\pm$3 & 726$\pm$11 & 241$\pm$24 & 976$\pm$10 & 416$\pm$3 \Tstrut\Bstrut\\
    V$_{rot}$\tablenotemark{a} [km s$^{-1}$] & 212$\pm$2 & 105$\pm$2 & 196$\pm$1 & 145$\pm$3 & 218$\pm$6 & 244 & 315$\pm$12 \Tstrut\Bstrut\\
    WSRT $f_{19}$ & 0.011$\pm$0.001 & 0.014$\pm$0.001 & --- & --- & 0.064$\pm$0.004 & 0.006$\pm$0.001 & 0.004$\pm$0.001 \Tstrut\Bstrut\\
    GBT $f_{19}$ & 0.005$\pm$0.001 & 0.008$\pm$0.001  & 0.005$\pm$0.001 & 0.003$\pm$0.001 & 0.046$\pm$0.002 & 0.008$\pm$0.001 & 0.003$\pm$0.001 \Tstrut\Bstrut\\
    \\[-1.0em]
    \hline
    \end{tabular}}
      \caption{Summary of Derived Properties}
    \tablenotetext{1}{Rotation velocity taken from \textit{HYPERLEDA} search}
\label{tab:derivedParams}
\end{table*}

We presented an initial analysis of deep ($N_{HI}$$\sim$10$^{18}$ cm$^{-2}$) GBT observations of four sources (NGC891, NGC925, NGC4414, and NGC4565) out of 24 total sources in the HALOGAS survey. These observations are among the most sensitive $\hi$ observations of external galaxies to date. In order to directly compare interferometer and single dish data, we solve for an optimal smoothing kernel specific to each source and convolve the WST data to GBT angular resolution. Our main conclusions are:

\begin{itemize}
\item  As we do not find significant spatially extended $\hi$ features, we conclude that the WSRT data do an excellent job recovering the diffuse (18 $\leq$ log$_{10}$($N_{HI}$) $\leq$ 19) $\hi$ around these four sources. In the case of NGC925, we detect about 20\% more $\hi$ than observations done with the VLA as part of the THINGS survey. The discrepancy is likely due in large part to the increased surface brightness sensitivity of the WSRT data since the ability to detect extended structure between the two surveys is very similar. The excellent agreement between the global $\hi$ profiles, cumulative $\hi$ mass as a function of $N_{HI}$, radial mean column density profiles, and radial cumulative flux for the GBT and convolved WSRT data provides additional evidence in support of this conclusion.

\item The cumulative $\hi$ mass as a function of $\hi$ column density reveals the diffuse $\hi$ associated with these galaxies does not change significantly over the range $log_10\left(N_{HI}/cm^{-2}\right)$ = 18.0 to $log_{10}\left(N_{HI}/cm^{-2}\right)$ = 19.5. The flat behavior is consistent with predictions from simulations, which show the neutral fraction is around 1\% at $log_10\left(N_{HI}/cm^{-2}\right)$ = 18.0. Scaling our GBT beam model to the peak column density of the GBT data and repeating our analysis to essentially simulate an unresolved source produces a similarly flat distribution, which suggests the lowest column density bins include some values that trace the extended structure of the GBT beam. That said, there is generally a moderate offset between the data and model cumulative HI mass distributions. The overall agreement between the GBT and WSRT data sets, if extended to the other sources in our survey, suggests future surveys must probe column densities at the $\sim$10$^{17}$ cm$^{-2}$ level to increase the probability of detecting $\hi$ associated with cosmic web structure or possibly cold-mode accretion.

\item We define a parameter, $f_{19}$, equal to one minus the ratio between the $\hi$ mass measured at and above log$_{10}$($N_{HI}$) = 19 and the source's total $\hi$ mass. We find that, on average (and excluding data that may suffer from resolution effects), this value is equal to 2\%, indicating the diffuse extended disks of these galaxies do not constitute a significant fraction of the overall mass.

\end{itemize} 

One observational method to differentiate between inflow/outflow origins is a measure of metallicity using UV absorption lines (e.g., S II) utilizing the Cosmic Origins Spectrograph on the $\textit{Hubble}$ Space Telescope. If a significant diffuse $\hi$ feature is seen around a source as we analyze the full survey, and a fortuitous background quasar along the line of sight, a metallicity of $\sim$ 0.1 $Z_{\odot}$ would be highly indicative of a CGM origin. 

To establish or rule out cold mode accretion as a feasible avenue for nearby galaxies to refuel their gas content, we must continue to analyze galaxies within the HALOGAS sample that satisfy the mass constraints set by simulations, show large diffuse $\hi$ mass fractions and low SFRs, and reside in low density environments. Due to our small sample size we can only present foundational work to uncover any underlying correlations between large mass fractions of diffuse $\hi$ and galaxy properties. Future work will focus on the analysis techniques presented here in order investigate the origins of these large $\hi$ filaments, apply short spacing corrections to the WSRT data, and continue the investigation into role of $\hi$ in galaxy evolution. 

\acknowledgments
 
We thank Richard Prestage and Jay Lockman for insightful discussions on the behavior and geometry of the innermost sidelobes of the GBT. We are also grateful to Filippo Fraternali for his helpful discussions on the caveats of convolving high-resolution data and interpretation of residual emission. Finally, we thank the anonymous referee for their constructive comments that helped to focus the analysis and conclusions presented in this work. This study was funded by the NSF CAREER grant AST-1149491. The National Radio Astronomy Observatory is a facility of the National Science Foundation operated under cooperative agreement by Associated Universities, Inc. T.~H.~J.~acknowledges financial support from the National Research Foundation (NRF; South Africa).
\bibliographystyle{apj}

\end{document}